\documentclass[
preprint,
tightenlines,
superscriptaddress,
nofootinbib,
 amsmath,amssymb,
 aps,
]{revtex4-1}

\usepackage{epsfig,amsmath,amssymb,verbatim,mathrsfs,hyperref}

\def\beq{\begin{eqnarray}}
\def\eeq{\end{eqnarray}}
\def\bea{\begin{eqnarray}}
\def\eea{\end{eqnarray}}

\def\tev{\, {\rm TeV}}
\def\gev{\, {\rm GeV}}
\def\mev{\, {\rm MeV}}

\newcommand{\gsim}{\lower.7ex\hbox{$\;\stackrel{\textstyle>}{\sim}\;$}}
\newcommand{\lsim}{\lower.7ex\hbox{$\;\stackrel{\textstyle<}{\sim}\;$}}

\def\mpl{M_{\rm Pl}}

\newcommand{\ccdot}{\!\cdot\!}
\newcommand{\nnmb}{\nonumber}
\newcommand{\del}{\partial}

\newcommand{\lrf}[2]{\left(\frac{#1}{#2}\right)}

\newcommand{\Mpl}{M_\mathrm{Pl}}
\newcommand{\TeV}{\mathrm{TeV}}
\newcommand{\GeV}{\mathrm{GeV}}
\newcommand{\MeV}{\mathrm{MeV}}
\newcommand{\TRH}{T_\mathrm{RH}}
\newcommand{\TRHx}{T_\mathrm{RH}^x}
\newcommand{\TBBN}{T_\mathrm{BBN}}
\newcommand{\Tfo}{T_\mathrm{fo}}
\newcommand{\Teq}{T_\mathrm{eq}}
\newcommand{\BR}{\mathrm{BR}}
\newcommand{\mphi}{m_{\varphi}}
\newcommand{\mgrav}{m_{3/2}}
\newcommand{\yt}{{\widetilde{Y}}}
\newcommand{\ytc}{{\widetilde{Y}^c}}
\newcommand{\msoft}{m_{\mathrm{soft}}}
\newcommand{\Neff}{N_{\mathrm{eff}}}

\begin{document}

\title{Confronting the Moduli-Induced LSP Problem}

\author{Nikita Blinov}
\email{nblinov@triumf.ca}
\affiliation{Theory Department, TRIUMF, 4004 Wesbrook Mall, Vancouver, BC V6T 2A3, Canada}
\affiliation{Department of Physics and Astronomy, University of British Columbia, Vancouver, BC V6T 1Z1, Canada}
\author{Jonathan Kozaczuk} 
\email{jkozaczuk@triumf.ca}
\affiliation{Theory Department, TRIUMF, 4004 Wesbrook Mall, Vancouver, BC V6T 2A3, Canada}
\author{Arjun Menon}
\email{aamenon@uoregon.edu}
\affiliation{Department of Physics, University of Oregon, Eugene, OR 97403, USA}
\author{David E. Morrissey}
\email{dmorri@triumf.ca}
\affiliation{Theory Department, TRIUMF, 4004 Wesbrook Mall, Vancouver, BC V6T 2A3, Canada}

\begin{abstract}
Moduli fields with Planck-suppressed couplings to light species 
are common in string compactifications. 
Decays of these moduli can reheat the universe at a late time
and produce dark matter non-thermally.  For generic moduli fields
motivated by string theory with masses similar to that of the gravitino and TeV-scale
superpartners in the minimal supersymmetric Standard Model~(MSSM), 
the non-thermal production of the lightest superpartner~(LSP) 
tends to create an unacceptably large relic density or too strong of an
indirect detection signal.  
We call this the moduli-induced LSP problem of the MSSM.  
In this paper we investigate extensions of the MSSM
containing new LSP candidates that can alleviate this tension.  
We examine the viability of this scenario in models with light 
Abelian and non-Abelian hidden sectors, and symmetric or 
asymmetric dark matter. 
In these extensions it is possible, though somewhat challenging, 
to avoid a moduli-induced LSP problem.  In all but the asymmetric 
scenario, the LSP can account for only a small fraction of the 
observed dark matter density.

\end{abstract}

\maketitle
\setcounter{page}{2}


\section{Introduction\label{sec:intro}}

  Moduli are light scalar fields with only higher-dimensional couplings to
other light species.  They arise frequently in theories with 
supersymmetry~\cite{Coughlan:1983ci,Ellis:1986zt,Banks:1993en}, 
and they appear to be a generic feature 
of string 
compactifications~\cite{Polchinski:1998rq,Polchinski:1998rr,Acharya:2012tw}.  
Moduli fields can also have significant implications 
for the cosmological history of the Universe,  
the mass spectrum of the supersymmetric partners of the Standard Model~(SM),
and the density of dark matter~(DM) 
today~\cite{de Carlos:1993jw,Quevedo:2002xw,Acharya:2008bk}.

  A modulus field can alter the standard cosmology if it is 
significantly displaced from the minimum of its potential in the
early Universe, as can occur following primordial inflation~\cite{Dine:1995uk}.
The modulus will be trapped by Hubble damping until $H \sim \mphi$,
at which point it will begin to oscillate.   The energy density of 
these oscillations dilutes in the same way as non-relativistic matter, 
and can easily come to dominate the expansion of 
the Universe.\footnote{Such an early matter-dominated phase 
might also leave an observable signal in gravitational 
waves at multiple frequencies~\cite{Durrer:2011bi}
or modify cosmological observables~\cite{Erickcek:2011us,Easther:2013nga, Iliesiu:2013rqa,Fan:2014zua}.} 
This will continue until the modulus decays at time 
$t\sim \Gamma_\varphi^{-1}$, transferring the remaining oscillation energy
into radiation.  At this point, called \emph{reheating}, 
the radiation temperature is approximately~\cite{Moroi:1999zb} 
\beq
\TRH ~\sim~ (5\,\MeV)\,\lrf{\mpl}{\Lambda}\lrf{\mphi}{100\;\TeV}^{3/2} \ ,
\label{eq:trhintro}
\eeq
where $\Lambda$ is the heavy mass scale characterizing the coupling
of the modulus to light matter.  To avoid disrupting primordial 
nucleosynthesis, the reheating temperature should be greater 
than about $\TRH \gtrsim 5\,\MeV$~\cite{Hannestad:2004px},
and this places a lower bound on the modulus mass.

  The mass of a modulus field is determined by its potential.
In string compactifications, multiple moduli typically appear
in the low-energy supergravity theory as components
of chiral multiplets with couplings to other fields suppressed
by powers of $\Lambda \sim \mpl$~\cite{Acharya:2012tw}.  
While the potentials of many of these moduli are not completely 
understood, certain features do seem to be fairly universal.
For example, moduli masses of $\mphi \sim \mgrav$ are expected 
when the potential arises mainly from 
supersymmetry breaking~\cite{Bose:2013fqa}.  
Moduli may also have supersymmetric potentials~\cite{Kachru:2003aw},
and $\mphi \gg \mgrav$ is found in some cases~\cite{Kallosh:2006dv,Fan:2011ua}.
However, $\mphi \sim \mgrav$ is still frequently obtained 
from supersymmetric potentials once the constraint of a very small 
vacuum energy is imposed~\cite{Fan:2011ua}.
Thus, a plausible generic expectation from string theory is that there
exists at least one modulus field with $\mphi \sim \mgrav$ 
and $\Lambda \sim \mpl$~\cite{Acharya:2012tw}.\footnote{
The LARGE Volume Scenario of Refs.~\cite{Balasubramanian:2005zx,Cicoli:2008va} 
is a notable exception to this.}  
Other heavier moduli may be present, but since the lightest 
and most weakly-coupled modulus will decay the latest, 
it is expected to have the greatest impact on the present-day cosmology. 

  Putting these two pieces together, acceptable reheating
from string moduli suggests $\mphi \sim \mgrav \gtrsim 100\,\tev$.
This has important implications for the masses of the SM superpartner fields.
Surveying the most popular mechanisms of supersymmetry
breaking mediation~\cite{Martin:1997ns}, the typical size
of the superpartner masses is
\beq
\msoft \sim \left\{
\begin{array}{ccccc}
\mgrav&&\text{gravity mediation}\\
\lrf{L \mpl}{M_*}\mgrav&&\text{gauge mediation}\\
L\mgrav&&\text{anomaly mediation}
\end{array}\right.
\eeq
where $L \sim g^2/(4\pi)^2$ is a typical loop factor and $M_* \ll \mpl/L$
is the mass of the gauge messengers.
Of these mechanisms, only anomaly mediation~(AMSB) allows for 
superpartners that are light enough to be directly observable
at the Large Hadron Collider~(LHC)~\cite{Randall:1998uk,Giudice:1998xp}.
Contributions to the soft terms of similar size can also be generated 
by the moduli themselves~\cite{Choi:2004sx,Choi:2005ge}, or other 
sources~\cite{Nelson:2002sa,Sundrum:2004un,Hsieh:2006ig,Everett:2008ey}.  
However, for these AMSB and AMSB-like contributions to be dominant,
the gravity-mediated contributions must be suppressed~\cite{Randall:1998uk},
which is non-trivial for the scalar soft 
masses~\cite{Luty:2001zv,Schmaltz:2006qs,Anisimov:2002az,Kachru:2006em,Kachru:2007xp}.
An interesting intermediate scenario is mini-split 
supersymmetry where the dominant scalar soft masses come from direct gravity
mediation with $\msoft \sim \mgrav$, while the gaugino soft
masses are AMSB-like~\cite{Wells:2004di,Hall:2011jd,Arvanitaki:2012ps,ArkaniHamed:2012gw,Ibe:2011aa,Ibe:2012hu}.

  Moduli reheating can also modify dark matter 
production~\cite{Moroi:1999zb,Gelmini:2006pw,Acharya:2009zt,
Arcadi:2011ev,Moroi:2013sla}.
A standard weakly-interacting massive particle~(WIMP) $\chi$ 
will undergo thermal freeze-out at temperatures near $\Tfo \sim m_{\chi}/20$.
If this is larger than the reheating temperature, the WIMP density
will be strongly diluted by the entropy generated from moduli decays.
On the other hand, DM can be created \emph{non-thermally} as moduli
decay products.  A compelling picture of non-thermal dark
matter arises very naturally for string-like moduli and an AMSB-like
superpartner mass spectrum~\cite{Moroi:1999zb}.  The lightest (viable) 
superpartner~(LSP) in this case tends to be a wino-like neutralino.
These annihilate too efficiently to give the observed relic density 
through thermal freeze-out~\cite{ArkaniHamed:2006mb,Cohen:2013ama,Fan:2013faa}.
However, with moduli domination and reheating, the wino LSP can be
created non-thermally in moduli decays, and the correct DM density
is obtained for $M_2\sim 200\,\gev$ and $\mphi\sim 3000\,\tev$.

  This scenario works precisely because the wino annihilation
cross section is larger than what is needed for thermal
freeze-out.  Unfortunately, such enhanced annihilation rates
are strongly constrained by gamma-ray observations of the galactic 
centre by Fermi and HESS, and the non-thermal wino is ruled out
even for very conservative assumptions about the DM profile in the inner galaxy 
(\emph{e.g.} cored isothermal)~\cite{Cohen:2013ama,Fan:2013faa}.
A wino-like LSP can be consistent with these bounds if it is only
a subleading component of the total DM density.  Using the
AMSB relation for $M_2$ in terms of $\mgrav$, this forces
$\mphi/\mgrav \gtrsim 100$, significantly greater than the 
generic expectation~\cite{Fan:2013faa}.  
The problem is even worse for other neutralino LSP species,
since these annihilate less efficiently and an even larger
value of $\mphi \gg \mgrav$ is needed to obtain an acceptable relic density.
Furthermore, $\mphi > 2\mgrav$ also allows the modulus field to
decay to pairs of gravitinos.  The width for this decay is typically
similar to the total width to SM 
superpartners~\cite{Endo:2006zj,Nakamura:2006uc,Asaka:2006bv,Dine:2006ii}.  
For $\mphi \gg \mgrav > 30\,\tev$, the gravitinos produced this way
will decay to particle-superpartner pairs before nucleosynthesis but after 
the modulus decays, and recreate the same LSP density problem 
that forced $\mphi \gg \mgrav$ in the first place.  

  These results suggest a degree of tension between reheating by 
string-motivated moduli (with $\mphi \sim \mgrav$ and
$\Lambda \sim \mpl$) and the existence of a stable TeV-scale LSP
in the minimal supersymmetric standard model~(MSSM).
This tension can be resolved if all relevant moduli have
properties that are slightly different from the na\"ive
expectation; for example $\mphi\gg \mgrav$ and 
$\BR(\varphi\to\psi_{3/2}\psi_{3/2}) \ll 1$~\cite{Dine:2006ii,Linde:2011ja},
an enhanced modulus decay rate with $\mphi\sim \mgrav$ 
and $\Lambda < \mpl$~\cite{Evans:2013nka},
or a suppressed modulus branching fraction 
into superpartners~\cite{Allahverdi:2013noa}.
Given the challenges and uncertainties associated with 
moduli stabilization in string theory, we focus on what seem
to be more generic moduli and we investigate a second approach:
extensions of the MSSM that contain new LSP candidates 
with smaller relic densities or that are more difficult to detect 
than their MSSM counterparts.

  In this paper we investigate extensions of the MSSM containing
additional hidden gauge sectors as a way to avoid the 
moduli-induced LSP problem of the MSSM.  
Such gauge extensions arise frequently 
in grand-unified theories~\cite{Slansky:1981yr}
and string compactifications~\cite{Blumenhagen:2005mu,Blumenhagen:2006ci}.
We assume that the dominant mediation of supersymmetry breaking 
to gauginos is proportional to the corresponding gauge coupling,  
as in anomaly or gauge mediation, allowing the hidden sector gauginos
to be lighter than those of the MSSM if the former have a smaller 
coupling constant~\cite{Feng:2008ya,Hooper:2008im,ArkaniHamed:2008qp}.  
We also focus on the case of a single light modulus field with 
$\mphi\sim \mgrav$ and $\Lambda \sim \mpl$,
although similar results are expected to hold for multiple moduli
or for reheating by gravitino decays. 

  This paper is organized as follows.  In Section~\ref{sec:reheat} 
we review moduli cosmology and the resulting non-thermal  production of LSPs.
Next, in Section~\ref{sec:mssm} we examine in more detail the
tension between moduli reheating and a stable MSSM LSP.
In the subsequent three sections we present three extensions of
the MSSM containing new LSP candidates and examine their abundances
and signals following moduli reheating.   
The first extension, discussed in Section~\ref{sec:u1sym}, comprises a minimal 
supersymmetric $U(1)_x$ hidden sector.  
We find that this setup allows for a hidden sector LSP with a relic density lower than that of the wino and which is small enough to evade the current bounds from indirect detection.
In Section~\ref{sec:u1asym} we extend the $U(1)_x$ 
hidden sector to include an asymmetric dark matter candidate and find that 
it is able to saturate the entire observed DM relic density while avoiding 
constraints from indirect detection. 
In Section~\ref{sec:sun} we investigate a pure non-Abelian hidden sector, 
and show that the corresponding gaugino LSP can provide an acceptable 
relic density and avoid constraints from indirect detection, although it is also very strongly
constrained by its effect on structure formation and the cosmic
microwave background.
Finally, Section~\ref{sec:conc} is reserved for our conclusions.

\section{Moduli Reheating and Dark Matter\label{sec:reheat}}

  In this section we review briefly the cosmology of moduli oscillation
and decay, as well as the associated non-thermal production of dark matter.

\subsection{Moduli Reheating}
  
  A modulus field $\varphi$ is very likely to develop a large
initial displacement from the minimum of its potential before or
during the course of primordial inflation~\cite{Coughlan:1983ci,Dine:1995uk}.  
Hubble damping will trap the modulus until $H\sim \mphi$, 
at which point it will start to oscillate coherently.  
For even moderate initial displacements,
these oscillations will eventually dominate over radiation.
The time evolution of the modulus oscillation energy density for $H < \mphi$
is given by
\beq
\dot{\rho}_\varphi + 3H\rho_\varphi + \Gamma_\varphi\rho_\varphi = 0 \ ,
\label{eq:boltzmod}
\eeq
where $\Gamma_\varphi$ is the modulus decay rate. 
For a modulus field with $\Mpl$-suppressed couplings 
\beq
\Gamma_\varphi = \frac{c}{4\pi}\frac{m_\varphi^3}{\Mpl^2} \ ,
\label{eq:moddecay}
\eeq 
where $c$ is a model-dependent number with a typical range 
of $10^{-3} < c < 100$~\cite{Fan:2013faa}.\footnote{
Values of $c$ much larger than this can be interpreted
as corresponding to a suppression scale $\Lambda < \mpl$.}
As the modulus oscillates, it decays to radiation with the radiation
density becoming dominant once more when $H \sim \Gamma_{\varphi}$. 

  The evolution of the radiation density $\rho_R$ follows from 
the First Law of thermodynamics: 
\beq
\frac{d\rho_R}{dt} + 3 H (\rho_R + p_R) = \Gamma_\varphi \rho_\varphi,
\label{eq:boltzrad}
\eeq
where $p_R$ is the radiation pressure. The right hand side is the rate 
of energy injection into the bath, of which moduli decays are assumed to be
the dominant source. Contributions from DM annihilation can also be included,
but these do not make much difference when the DM is lighter than 
the modulus field.  The radiation density is used to define 
the temperature through 
\beq
\rho_R = \frac{\pi^2}{30}g_*(T)T^4,
\eeq
where $g_*(T)$ is the effective number of relativistic degrees 
of freedom~\cite{Kolb:1990vq}.  Reheating is said to occur when 
radiation becomes the dominant energy component of the Universe,
corresponding to $H(\TRH)\simeq \Gamma_{\varphi}$. 
Following Refs.~\cite{Moroi:1999zb},
we define the reheating temperature $\TRH$ to be:
\beq
\TRH &=& \left(\frac{90}{\pi^2 g_*(\TRH)} \right)^{1/4} \sqrt{\Gamma_\varphi \Mpl}
\label{eq:trh}\\
&\simeq& 
(5.6\;\MeV)\; c^{1/2}\left(\frac{10.75}{g_*}\right)^{1/4}
\left(\frac{\mphi}{100\;\TeV}\right)^{3/2} . \nnmb
\eeq 
 Here $\Mpl\simeq 2.4\times 10^{18}\gev$ is the reduced Planck mass. 
The reheating temperature
 $\TRH$ should exceed $5\,\MeV$ to preserve the predictions 
of primordial nucleosynthesis~\cite{Hannestad:2004px}.\footnote{
We have adjusted for our slightly different definition
of $\TRH$ relative to Ref.~\cite{Hannestad:2004px} in the quoted bound.} 
For $c = 1$ this implies that $m_\varphi \gtrsim 100\;\TeV$.

\subsection{Non-Thermal Dark Matter}

    Moduli decays can also produce stable massive particles, such as
a self-conjugate dark matter candidate $\chi$~\cite{Moroi:1999zb}.
This is described by
\beq
\frac{dn_\chi}{dt} + 3 H n_\chi = 
\frac{\mathcal{N}_\chi \Gamma_\varphi}{m_\varphi} \rho_\varphi
- \langle \sigma v\rangle(n_\chi^2 - n_{\mathrm{eq}}^2),
\label{eq:boltzdm}
\eeq
where 
$\langle\sigma v\rangle$ is the thermally averaged annihilation cross-section, 
$n_\mathrm{eq} = g T m_\chi ^2 K_2(m_\chi/T)/2\pi^2$ is the 
equilibrium number density, with $g$ being the number of internal degrees of freedom
and $\mathcal{N}_\chi$ is the average number of $\chi$ particles
produced per modulus decay.\footnote{This includes $\chi$ produced
in direct decays, as well as rescattering~\cite{Harigaya:2014waa} 
and decay cascades.} 
Values of $\mathcal{N}_{\chi}\sim 1$ are usually expected 
when $\chi$ is the LSP~\cite{Bose:2013fqa,Kaplan:2006vm}.
Together, Eqs.~(\ref{eq:boltzmod}, \ref{eq:boltzrad}, 
\ref{eq:boltzdm}) and the Friedmann equation form a closed set of
equations for the system.

  The general solution of these equations interpolates between three
distinct limits~\cite{Moroi:1999zb,Gelmini:2006pw,Arcadi:2011ev,Pallis:2004yy}.  
For reheating temperatures above the thermal freeze out
temperature $\Tfo$ of $\chi$, the final $\chi$ density approaches
the thermal value.  When $\TRH < \Tfo$, annihilation may or may not
be significant depending on $\langle\sigma v\rangle$ 
and $\mathcal{N}_{\chi}$.  
Smaller values imply negligible $\chi$ annihilation after reheating 
and a final relic density of about~\cite{Arcadi:2011ev}
\beq
\Omega_{\chi}h^2 &\simeq& 
\frac{3}{4}\mathcal{N}_{\chi}\lrf{m_{\chi}}{\mphi}\TRH\lrf{s_0}{\!\rho_c/h^2\!}
\label{eq:noann}\\
&\simeq& (1100)\,\mathcal{N}_\chi
\left(\frac{m_\chi}{100\,\gev}\right)
\left(\frac{\TRH}{5\,\mev}\right)
\left(\frac{100\,\TeV}{\mphi}\right) \ ,
\nnmb
\eeq
where $s_0$ is the entropy density today and $\rho_c/h^2$ is 
the critical density.  
Larger values of $\mathcal{N}_{\chi}$ or $\langle\sigma v\rangle$
lead to significant annihilation among the $\chi$ during the
reheating process, giving a relic density of~\cite{Acharya:2009zt,Arcadi:2011ev}
\beq
\Omega_{\chi}h^2 &\simeq& 
\frac{m_{\chi}\Gamma_{\varphi}}{\langle\sigma v\rangle s_{\rm RH}}
\lrf{s_0}{\!\rho_c/h^2\!}
\label{eq:reann}\\
&\simeq& (0.2)
\lrf{m_\chi/20}{\TRH}
\lrf{3\times 10^{-26}\mathrm{cm}^3/\mathrm{s}}{\langle\sigma v\rangle}
\lrf{10.75}{g_*}^{1/2} \nnmb\\
&\simeq& 
(200)\,c^{-1/2}
\lrf{m_{\chi}}{100\,\gev}
\lrf{3\times 10^{-26}\text{cm}^3/\text{s}}{\langle\sigma v\rangle}
\lrf{100\,\tev}{\mphi}^{3/2}\lrf{10.75}{g_*}^{1/4}
\ . \nnmb
\eeq
We emphasize that the expressions of Eqs.~(\ref{eq:noann},\ref{eq:reann}) 
are only approximations valid to within a factor of order unity.
In what follows we solve this system numerically using the methods of 
Refs.~\cite{Chung:1998rq,Giudice:2000ex}.  For $\TRH < \Tfo$
and $\mathcal{N}_{\chi}$ not too small, the reannihilation 
scenario is usually the relevant one~\cite{Arcadi:2011ev}.

\subsection{Scaling Relations}

  It is instructive to look at how the relation of Eq.~\eqref{eq:reann}   
scales with the relevant couplings and masses~\cite{Feng:2008ya}.
Motivated by the MSSM wino in anomaly
mediation, we will assume that the dark matter mass scales with 
a coupling $g_{\chi}$ according to
\beq
m_{\chi} = r_{\chi}\frac{g_{\chi}^2}{(4\pi)^2}m_{3/2} \ ,
\label{eq:mscale}
\eeq
for some parameter $r_{\chi}$.
We will assume further that the dark matter annihilation cross section
scales with the coupling as well,
\beq
\langle\sigma v\rangle = \frac{k_{\chi}}{4\pi}\frac{g_{\chi}^4}{m_{\chi}^2} \ ,
\label{eq:sigscale}
\eeq
for some parameter $k_{\chi}$.  For an AMSB-like wino, 
the $r$ and $k$ parameters are~\cite{Moroi:1999zb}
\beq
r_{2} &\simeq& 1 \ , \\
k_{2} &\simeq& 
2\,\frac{~~[1-(m_W/M_2)^2]^{3/2}}{[2-(m_W/M_2)^2]^2}
~~\to~~ 1/2 \ ,
\eeq
with $g_{\chi} = g_2 \simeq 0.65$, and the last expression
neglects coannihilation with charginos, which can be suppressed
at low reheating temperatures~\cite{Arcadi:2011ev}.

  With these assumptions, the thermal $\chi$ abundance is
\beq
\Omega_{\chi}^{\mathrm{th}}h^2 &\simeq& (5.5\times 10^{-3})\,
\frac{r_{\chi}^2}{k_{\chi}}\lrf{m_{\chi}/\Tfo}{20}
\lrf{\mgrav}{100\,\tev}^2\lrf{106.75}{g_*}^{1/2}
\label{eq:thscale}
\eeq
independent of the specific mass or coupling.  
This is no longer true of non-thermal DM produced by moduli decays, 
where the mass dependence is different.  Rewriting Eq.~\eqref{eq:reann} subject
to the assumptions of Eqs.~(\ref{eq:mscale},\ref{eq:sigscale}),
we obtain
\beq
\Omega_{\chi}h^2 ~\simeq~ 
15\,c^{-1/2}\,\lrf{r_{\chi}^3/k_{\chi}}{r_2^3/k_2}\,\lrf{g_{\chi}}{g_2}^2\lrf{m_{3/2}}{\mphi}^{3}\lrf{\mphi}{100\,\tev}^{3/2}\lrf{10.75}{g_*}^{1/4} \ .
\label{eq:omscale}
\eeq
This result shows that reducing the coupling or the modulus mass
suppresses the non-thermal relic density.  It also makes clear that 
$\mphi > \mgrav$ is needed to obtain an acceptable wino abundance
within the reannihilation regime.

\subsection{Gravitino Production and Decay}
 
  Our previous discussion of moduli reheating did not take 
gravitinos into account.  Moduli can also decay to gravitinos 
if $\mphi > 2\mgrav$, and the corresponding branching ratio
$\text{BR}_{3/2}$ is expected to be on the order of unity 
unless some additional structure is 
present~\cite{Endo:2006zj,Nakamura:2006uc,Asaka:2006bv,Dine:2006ii}.  
For $\mphi \sim 2\mgrav$,
the gravitinos will decay at about the same time as the moduli
and our previous results for the moduli-only case are expected
to apply here as well.  On the other hand, if $\mphi \gg \mgrav$
and $\text{BR}_{3/2}$ is not too small, the gravitinos produced
by decaying moduli are likely to come to dominate the energy
density of the Universe before they themselves decay.
We examine this possibility here, and show that our results
for moduli decay can be applied to this scenario as well 
after a simple reinterpretation of parameters.

  If the gravitino is not the LSP, it will decay to lighter 
particle-superpartner pairs with 
\beq
\Gamma_{3/2} &=& \frac{d}{4\pi}\frac{\mgrav^3}{\mpl^2} \ ,
\eeq
where $d = 193/96$ if all MSSM final states are open
and $d= (1+3+8)/8=3/2$ if only gaugino modes are available~\cite{Moroi:1995fs}.
These decays will not appreciably disrupt BBN for 
$m_{3/2} \gtrsim 30\,\tev$, but they can produce a significant 
amount of LSP dark matter.

  For $\mphi \gg \mgrav$, the modulus will decay much earlier
than the gravitino (unless $c\ll d$).  The gravitinos produced
by moduli decays at time $t_i \simeq \Gamma_{\varphi}^{-1}$ 
will be initially relativistic with $p/\mgrav=\mphi/2\mgrav$.
Their momentum will redshift with the expansion of the Universe,
and they will become non-relativistic at time
\beq
t_{\rm nr} \simeq \frac{d}{4c}\lrf{\mgrav}{\mphi}\Gamma_{3/2}^{-1} \ ,
\eeq
where we have assumed that the Universe is radiation-dominated
after moduli reheating.  Thus, the gravitinos produced in moduli decays 
become non-relativistic long before they decay for $\mgrav/\mphi\ll 1$ 
(and $c\sim d$). 
While $t_{\rm nr} < t < \Gamma_{3/2}^{-1}$, the gravitinos will behave like matter.
The quantity $\mgrav n_{3/2}$ begins to exceed the (non-gravitino)
radiation density at time
\beq
t \simeq \frac{d}{c}\lrf{1-\text{BR}_{3/2}}{\text{BR}_{3/2}}^2
\lrf{\mgrav}{\mphi}^3\Gamma_{3/2}^{-1}
\eeq
Again, this is much earlier than the gravitino decay time
for $\mgrav/\mphi \ll 1$ unless $\text{BR}_{3/2}$ or $c/d$
is suppressed.\footnote{We have assumed radiation domination
here, but a similar result holds for matter domination.}
    
  The scenario that emerges for $\mgrav \ll \mphi$, $c\sim d$,
and $\text{BR}_{3/2}\sim 1$ is very similar to a second stage of
moduli reheating: the gravitinos produced in moduli decays become 
non-relativistic and come to dominate the energy density of 
the Universe until they decay at time $t\simeq \Gamma_{3/2}^{-1}$, 
at which point they reheat the Universe again.  
Dark matter will also be created by
the gravitino decays, with at least one LSP produced per decay
(assuming $R$-parity conservation).  The large gravitino density
from moduli decays can interfere with nucleosynthesis or produce
too much dark matter, and is sometimes called
the moduli-induced gravitino 
problem~\cite{Endo:2006zj,Nakamura:2006uc,Asaka:2006bv,Dine:2006ii}.  

  We will not discuss gravitinos much for the remainder of this paper.
Instead, we will focus mainly on the case of $\mphi\sim \mgrav$, where
the presence of gravitinos does not appreciably change our 
results~\cite{Arcadi:2011ev}.
However, our findings can also be applied to scenarios with 
$\mphi\gg \mgrav$, $c\sim d$, and $\text{BR}_{3/2} \sim 1$
with the moduli decays reinterpreted as gravitino decays
(\textit{i.e.} $\mphi \to \mgrav$, $c\to d$, $\mathcal{N}_{\chi}\to 1$).

\section{Moduli Reheating and the MSSM\label{sec:mssm}}

    The discussion of Section~\ref{sec:reheat} shows that the LSP relic
density is enhanced in the moduli-decay scenario relative to thermal
freeze out unless the fraction of decays producing LSPs $\mathcal{N}_\chi$
is very small.  In Ref.~\cite{Fan:2013faa}, this observation was used 
to put a very strong constraint on wino-like LSPs produced by moduli decays.  
In this section we apply these results to more general MSSM 
neutralino LSPs, and we argue that the MSSM has a 
\emph{moduli-induced LSP problem} for $\mphi\sim\mgrav$,
$c\sim 1$, and $\mathcal{N}_{\chi}$ not too small.  See also 
Refs.~\cite{Allahverdi:2012wb,Roszkowski:2014lga} for related analyses.

  Consider first a wino-like LSP with an AMSB-like mass.  
Direct searches at the LHC imply that the mass must lie
above $m_{\chi_1^0} \gtrsim 270\,\gev$ if it is nearly 
pure wino~\cite{Aad:2013yna},
although smaller masses down to the LEP limit 
$m_{\chi_1^{\pm}} \gtrsim 104\,\gev$
are possible if it has moderate mixing with a Higgsino~\cite{lepchargino}.
Examining Eq.~\eqref{eq:omscale}, the moduli-induced wino relic density
(in the reannihilation regime) tends to be larger than the observed
DM density, and indirect detection places an even stronger bound
of $\Omega_{\chi}h^2 \lesssim 0.05$~\cite{Fan:2013faa}.
Fixing $m_{\chi} = 270\,\gev$, a relic density of this size
can be obtained with the very optimistic combination of parameter values 
$c=100$, $\mphi=2\mgrav$, and $r_{\chi}/r_2 \lesssim 0.3$.  
Such a reduction in $r_{\chi}/r_2$ can arise from supersymmetry-breaking
threshold corrections~\cite{Gherghetta:1999sw,Gupta:2012gu} or moduli-induced
effects~\cite{Choi:2004sx,Choi:2005ge}, but requires a significant accidental
cancellation relative to the already-small 
AMSB value of $r_2$~\cite{Fan:2013faa}.

  A small effective value of $r_{\chi}< r_2$ could also arise 
from $|\mu| \ll |M_2|$ and a corresponding Higgsino-like LSP.  
The reduction in the relic density in this case is countered
by a smaller annihilation cross section: for $\mu\gg m_W$, 
heavy scalars, and neglecting coannihilation
we have $g_{\chi}\simeq g_2$, $r_{\chi}=(\mu/M_2)$, and
$k_{\chi} \simeq (3+2t_W^2+t_W^4)/128 \simeq 0.03$~\cite{ArkaniHamed:2006mb} 
(where $t_W \equiv \tan\theta_W$, with $\theta_W$ the Weinberg angle).
To investigate this possibility in more detail, 
we set $\mphi/\mgrav = 1,\,10,\,100$ and $c=1$, 
and compute the moduli-induced LSP relic density for various values 
of $\mu/M_2$ and $\mgrav$.  
In doing so, we fix $M_2$ to its AMSB value with $c=1$ and $\mphi = \mgrav$,
and we compute the annihilation cross section in 
\texttt{DarkSUSY}~\cite{Gondolo:2004sc,darksusy}.  
For the other MSSM parameters, we set $\tan\beta=10$, $m_A=1000\,\gev$,
$\tilde{m}=2000\,\gev$ for all scalars, and we fix $A_t$ such that
$m_h = 126 \pm 1\,\gev$.

\begin{figure}[ttt]
\centering
\includegraphics[width=17cm]{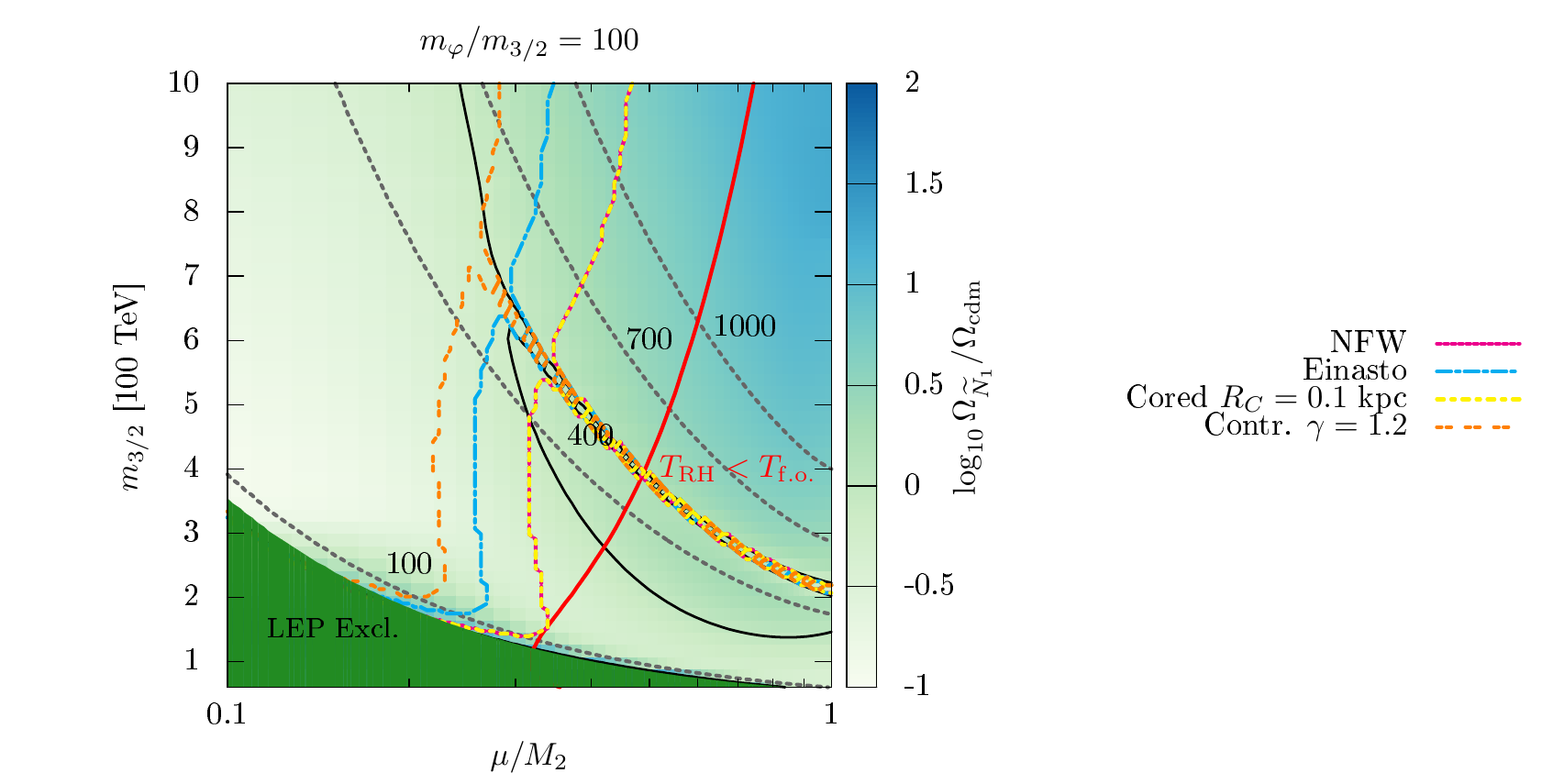}
\caption{
Relic density and constraints from indirect detection~(ID) 
for a mixed Higgsino-wino LSP produced by moduli reheating
as a function of $\mu/M_2$ and $m_{3/2}$.  The modulus parameters are
taken to be $m_\varphi/m_{3/2} = 100$, $c=1$, and $\mathcal{N}_{\chi}=1$.  
Contours of the LSP mass in GeV are given by the dashed grey lines.
The solid black contours show where $\Omega_{\chi_1^0} = \Omega_{\mathrm{cdm}}$.
The solid red line shows where $\TRH = \Tfo$: to the left of it 
we have $\TRH > \Tfo$; to the right $\TRH < \Tfo$ and the production is
non-thermal.  The remaining lines correspond to bounds from ID for 
different galactic DM distributions, and the area below and to the
right of these lines is excluded.
\label{fig:higgsino100}}
\end{figure}

  Fixing $\mphi/\mgrav=1,\,10$ we find no Higgsino-like points with
$\Omega_{\chi}h^2 \leq 0.12$, \emph{i.e.} for values which would appear to be generically expected from string theory.\footnote{
We also fail to find any such points for $c=100$ and $\mphi = 2\mgrav$.}
Smaller relic densities are found for $\mphi/\mgrav = 100$, 
and the results of our scan for this ratio are shown 
in Fig.~\ref{fig:higgsino100}.  The LSP relic density is smaller
than the total DM density to the left and below the solid black
line, while the grey dashed contours show the LSP mass. 
To the right of the red line, the reheating temperature lies above
the freeze-out temperature and the resulting density is thermal.
The coloured dashed contours correspond to 
bounds from indirect detection for different DM density profiles, 
excluding everything below and to the right of them.\footnote{The details
of our indirect detection analysis will be presented in the next section.}  
This figure also shows a funnel region with very low relic density
along the $m_{\chi_1^0} = 500\gev$ contour corresponding
to an $s$-channel $A^0$ pseudoscalar resonance. 

In general, for $\mphi\sim\mgrav$, we find that a Higgsino-like LSP 
also tends to produce too much dark matter when it is created in moduli 
reheating.  As for the wino, this can be avoided for larger values 
of $\mphi/\mgrav$, as demonstrated by Fig.~\ref{fig:higgsino100},
although one must still ensure that the very heavy modulus does not decay
significantly to gravitinos.

  These results can be extended to an arbitrary MSSM neutralino LSP. 
In general, mixing with a bino will further suppress the annihilation cross
section, leading to an overproduction of dark matter for $\mphi \sim \mgrav$.
The only loophole we can see is a very strong enhancement of the annihilation
from a resonance or coannihilation~\cite{Griest:1990kh}.  
This requires a very close mass
degeneracy either between $2m_{\chi}$ and the mass of the resonant state, 
or between $m_{\chi}$ and the coannihilating state, with coannihilation 
further suppressed at low reheating temperatures.
The only other viable LSP candidates in the MSSM are the sneutrinos.
These annihilate about as efficiently as a Higgsino-like LSP~\cite{Falk:1994es},
and therefore also tend to be overproduced.  The MSSM sneutrinos also
have a very large scattering cross section with nuclei, and bounds from
direct detection permit them to be only a small fraction 
of the total DM density~\cite{Srednicki:1986vj}.

  Having expanded slightly on the findings of Ref.~\cite{Fan:2013faa}, 
we conclude that a neutral MSSM LSP is typically overproduced in moduli 
reheating unless $\mphi \gg \mgrav$ (with tiny $\text{BR}_{3/2}$), 
$\mathcal{N}_{\chi}\ll 1$, or the decay coefficient $c \gg 100$ 
is very large.  None of these features appears to be generic 
in string compactifications.  We call this the 
\emph{moduli-induced MSSM LSP problem}.  
For this reason, we turn next to extensions of the MSSM with
more general LSP candidates that can potentially avoid this problem.

\section{Variation \#1: Hidden $U(1)$\label{sec:u1sym}}

  The first extension of the MSSM that we consider consists
of a hidden $U(1)_x$ vector multiplet $X$ and a pair of hidden 
chiral multiplets $H$ and $H'$ with charges $x_{H,H'} = \pm 1$.  
Motivated by the scaling relation of Eq.~\eqref{eq:omscale},
we take the characteristic gauge coupling and mass scale of the 
hidden sector to be significantly less than electroweak, 
along the lines of Refs.~\cite{Morrissey:2009ur,Feng:2011uf,Feng:2011in}.  
The LSP of the extended theory will therefore be the lightest hidden neutralino.
We also assume that the only low-energy interaction between the 
hidden and visible sectors is gauge kinetic mixing.  
Among other things, this allows the lightest MSSM superpartner
to decay to the hidden sector.
In this section we the investigate the contribution of the hidden LSP 
to the dark matter density following moduli reheating as well 
as the corresponding bounds from indirect and direct detection.

\subsection{Setup and Spectrum}

  The hidden superpotential is taken to be  
\begin{equation}
W_\mathrm{HS} = W_\mathrm{MSSM} - \mu' H H'\ ,
\label{eq:hidw}
\end{equation}
and the soft supersymmetry breaking terms are
\beq
-\mathscr{L}_\mathrm{soft} \supset m_H^2 |H|^2 + m_{H'}^2 |H'|^2 
+\left(-b' H H' +\frac{1}{2}M_x \widetilde{X}\widetilde{X} 
+\; \mathrm{h.c.}\right) \ .
\label{eq:hidsoft}
\eeq 
The only interaction with the MSSM comes from supersymmetric
gauge kinetic mixing in the form 
\beq
\mathscr L \supset \int\!d^2\theta\;\frac{\epsilon}{2}X^\alpha B_\alpha,  
\eeq
where $X$ and $B$ are the $U(1)_x$ and $U(1)_Y$ field strength superfields, 
respectively. 

  We assume that the gaugino mass is given by its AMSB value~\cite{Feng:2011uf}
\beq
M_x = b_x\frac{g_x^2}{(4\pi)^2}m_{3/2} \ ,
\eeq
where $b_x = 2$ and $g_x$ is the $U(1)_x$ gauge coupling.
Since pure AMSB does not provide a viable scalar spectrum
in the MSSM, we do not impose AMSB values on the scalar
soft terms in the hidden sector.  However, we do assume that
they (and $\mu'$) are of similar magnitude to their AMSB values, on the order 
of $(g_x^2/16\pi^2)m_{3/2}$.  This could arise if
the dynamics that leads to a viable MSSM spectrum also operates
in the hidden sector and that its effects are proportional
to the corresponding gauge coupling.    

  For a range of values of $\mu'$ and the soft terms,
the scalar components of $H$ and $H'$ will develop vacuum expectation
values, 
\beq
\langle H \rangle = \eta \sin \zeta,~\langle H' \rangle = \eta \cos \zeta \ .
\eeq
Correspondingly, the hidden vector boson $X_\mu$ receives a mass 
\beq
m_x = \sqrt{2} g_x \eta \ . 
\eeq 
The scalar mass eigenstates after $U(1)_x$ breaking consist of
two CP-even states $h_{1,2}^x$ (with $h_1^x$ the lighter of the two) and 
the CP-odd state $A^x$.  The fermionic mass eigenstates are
mixtures of the hidden Higgsinos and the $U(1)_x$ gaugino, 
and we label them in order of increasing mass as $\chi_{1,2,3}^x$.
Full mass matrices for all these states can be found 
in Refs.~\cite{Chan:2011aa,Morrissey:2014yma}.

\subsection{Decays to and from the Hidden Sector}

  Kinetic mixing allows the lightest MSSM neutralino to decay to
the hidden sector.  It can also induce some of the hidden states
to decay back to the SM.  We discuss the relevant decay modes here.

The MSSM neutralinos connect to the hidden sector through the bino. 
For AMSB gaugino masses, the bino soft mass is significantly heavier
than that of the wino, and the lightest neutralino $\chi_1^0$ tends to be 
nearly pure wino.  Even so, it will have a small bino admixture
given by the mass mixing matrix element $\mathbf{N}_{11}$. 
In the wino limit, it can be approximated by~\cite{ArkaniHamed:2006mb}
\beq
|\mathbf{N}_{11}| = 
\frac{c_W s_W m_Z^2 (M_2 + \sin 2\beta \mu)}{(M_1-M_2)(\mu^2-M_2^2)} \ .
\label{eq:winobinomix}
\eeq
With this mixing, the lightest MSSM neutralino will decay to
the hidden sector through the channels $\chi_1^0 \to \chi^x_k + S^x$, 
where $\chi^x_k$ are the hidden neutralinos and $S^x = h^x_{1,2},\;A^x,X_\mu$ 
are the hidden bosons, with total width~\cite{Chan:2011aa}
\begin{align}
\Gamma_{\chi_1^0}  & =  
\frac{\epsilon^2 g_x^2 |\mathbf{N}_{11}|^2}{4\pi}m_{\chi_1^0}\\
  & = (1.3\times 10^{-16}\;\mathrm{sec})^{-1} |\mathbf{N}_{11}|^2\left(\frac{\epsilon}{10^{-4}}\right)^2
\left(\frac{g_x}{0.1}\right)^2
\left(\frac{m_{\chi_1^0}}{100\;\GeV}\right) \ .\nnmb
\end{align}
The corresponding $\chi_1^0$ lifetime should be less than about
$\tau \lesssim 0.1\,\text{s}$ to avoid disrupting nucleosynthesis.
This occurs readily for MSSM gaugino masses below the TeV scale
and $\epsilon$ not too small.

  In the hidden sector, the $\chi_1^x$ neutralino will be stable
while the other states will ultimately decay to it or to the SM.
To ensure that $\chi_1^x$ is able to annihilate efficiently, it should also
be heavier than the vector $X^{\mu}$.  This implies that the hidden vector
will decay to the SM through kinetic mixing,
or via $X\to h_1^xA^x$.  For $m_x > 2m_e$, the vector decay width to the
SM is
\beq
\Gamma(X\to \mathrm{SM}+\mathrm{SM}) = R'\frac{\alpha\epsilon^2m_x}{3} \ ,
\eeq
where $R'$ is a constant on the order of unity that depends on 
the number of available final states.  This decay is much faster
than $\tau = 0.1\,\text{s}$ for $\epsilon \gtrsim 4\times 10^{-10}$ 
and $m_x \gtrsim 2m_{\mu}$.

  Of the remaining hidden states, the longest-lived is typically
the lightest CP-even scalar $h_1^x$.  The structure of the hidden
sector mirrors that of the MSSM, and this scalar is always lighter
than the vector at tree level.  Loop corrections are not expected to
change this at weak coupling.  As a result, the $h_1^x$ decays exclusively
to the SM through mixing with the MSSM Higgs scalars (via a Higgs portal
coupling induced by gauge kinetic mixing) or through 
a vector loop~\cite{Chan:2011aa}.
This decay is typically faster than $\tau = 0.1\,\text{s}$ for
$\epsilon \gtrsim 2\times 10^{-4}$ 
and $m_{h_1^x} \gtrsim 2m_{\mu}$~\cite{Morrissey:2014yma}.  

  Light hidden sectors of this variety are strongly constrained by
fixed-target and precision experiments~\cite{Bjorken:2009mm,Essig:2013lka}.  
For dominant vector decays to the SM, the strongest limits for $m_x > 2m_{\mu}$
come from the recent BaBar dark photon search~\cite{Lees:2014xha}, 
and limit $\epsilon \lesssim 5\times 10^{-4}$.   
As the vector mass approaches $m_x=20\,\mev$, fixed-target searches 
become relevant and constrain the mixing $\epsilon$ to extremely 
small values~\cite{Bjorken:2009mm,Essig:2013lka}. 
In this analysis, we will typically choose $m_{x} > 20\,\mev$ 
and $\epsilon \sim 10^{-4}$ so that the hidden sector is consistent 
with existing searches.

\subsection{Hidden Dark Matter from Moduli}

  Moduli decays are expected to produce both visible and hidden
particles and reheat both sectors.   The superpartners created by moduli 
decays will all eventually cascade down to the hidden neutralino LSP.
Kinetic mixing can allow the hidden LSP to thermalize by scattering
elastically with the SM background through the exchange of $X$ vector bosons.
The rate of kinetic equilibration depends on the typical energy at which 
the LSP is created, the reheating temperature, and the mass 
and couplings in the hidden sector~\cite{Arcadi:2011ev}.  
For optimistic parameter values we find that it is faster 
than the Hubble rate for $\TRH \gtrsim 5\,\mev$,
and we will assume here that such thermalization occurs.

If the net rate of superpartner production in moduli decays
is unsuppressed and the $\chi_1^x$ annihilation cross section is moderate, 
the $\chi_1^x$ LSPs will undergo additional annihilation to produce
a final relic density as described in Eq.~\eqref{eq:reann}.
The relevant annihilation modes of the LSP
are $\chi_1^x\chi_1^x \to h^x_1 h^x_1,\;Xh^x_1,\;XX$.  
Computing the corresponding annihilation rates using the method 
of Ref.~\cite{Gondolo:1990dk} near $T\sim\TRH$, 
we find that the $XX$ final state typically dominates provided it is
open, as we will assume here.   
Using these rates, we compute the relic abundance
of $\chi_1^x$ by numerically solving the system of equations
presented in Section~\ref{sec:reheat}.  In doing so, the decays of the
MSSM LSP and all hidden states are treated as being prompt.

Before presenting our numerical results, it is instructive to
examine the parametric dependence of the approximate solution
of Eq.~\eqref{eq:reann}.  Writing
\beq
\mu' = \xi\,M_x \ ,
\eeq
and focusing on a hidden Higgsino-like LSP with $\xi \leq 1$,  
we obtain $g_{\chi} =g_x$ and $r_\chi = 2\xi$ in
Eq.~\eqref{eq:omscale}.  Thus, smaller values of $\xi$ and $g_x$ 
are expected to produce decreased $\chi_1^x$ relic abundances.

\begin{figure}[ttt]
\centering
\includegraphics[width=8cm]{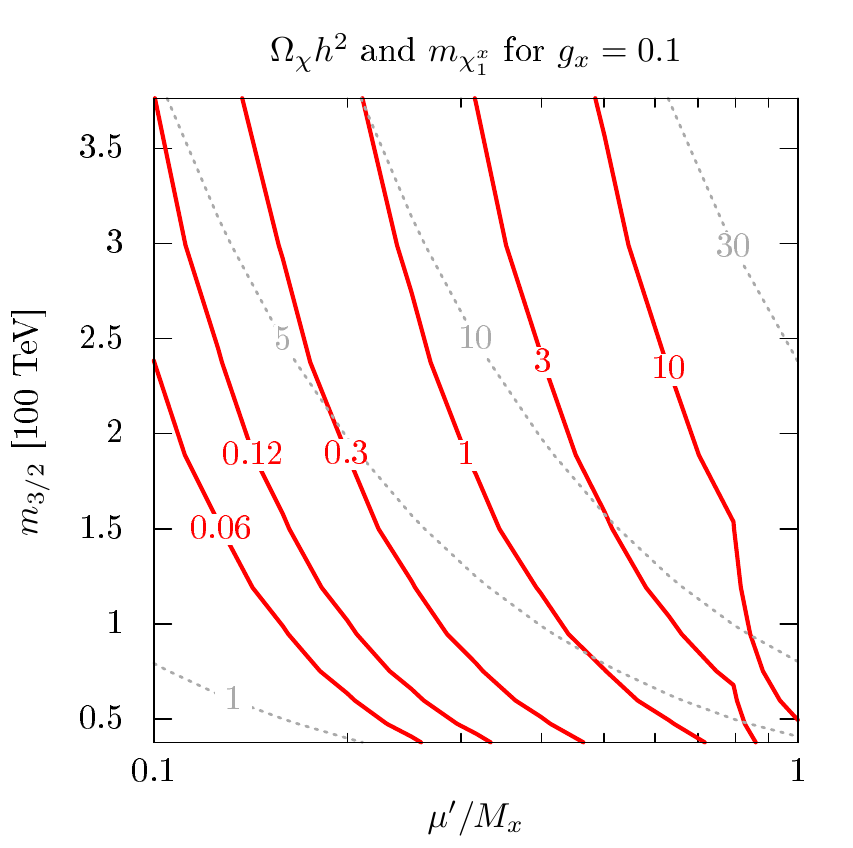}
\caption{Contours of the hidden neutralino $\chi_1^x$
mass in GeV (dashed grey) and moduli-generated relic abundance 
$\Omega_{\chi}h^2$ (solid red) as a function of of $\mu'/M_x$ and $\mgrav$.  
The moduli parameters are taken to be $\mphi = \mgrav$,
$c=1$, and $\mathcal{N}_{\chi}=1$, with the hidden-sector parameters
as described in the text.
\label{fig:hs_amsb}}
\end{figure}

  The results of a full numerical analysis are illustrated 
in Fig.~\ref{fig:hs_amsb}, where we show the contours of the 
final $\chi_1^x$ abundance (solid red) and DM mass (dashed grey) in the 
$\xi - m_{3/2}$ plane for $\mphi = \mgrav$, $c=1$, and $\mathcal{N}_{\chi}=1$. 
The range of $m_{3/2}$ considered corresponds to $M_2\in [100,1000]\;\GeV$, 
and the hidden sector parameters are taken to be
 $g_x = 0.1$, $\tan\zeta = 10$, and $m_x = 0.2\;\GeV$, $m_{A^x} = 10\;\GeV$.
The shape of the abundance contours in Fig.~\ref{fig:hs_amsb} 
is in agreement with the scaling predicted by Eq.~\eqref{eq:omscale}. 
We also see that $\xi =\mu'/M_x < 1$ is typically needed to avoid creating
too  much dark matter, and this implies some degree of fine tuning for hidden-sector
symmetry breaking.  Larger values of $\xi$ are allowed when
the moduli decay parameter $c$ is greater than unity, since this leads
to a higher reheating temperature and more efficient reannihilation.

\subsection{Constraints from Indirect Detection}
\label{subsec:id}

  While this extension of the MSSM can yield an acceptable
hidden neutralino relic density from moduli reheating, 
it is also constrained by indirect detection~(ID) searches for DM.\footnote{
Constraints from direct detection are not relevant; the $\chi_1^x$ LSP
is a Majorana fermion, and scatters off nuclei mainly through
a suppressed Higgs mixing coupling~\cite{Morrissey:2014yma}.}
The pair annihilation of hidden neutralinos can produce continuum 
photons at tree level from cascades induced by 
$\chi_1^x\chi_1^x \rightarrow XX$ with $X\rightarrow f \bar{f}$, 
as well as photon lines at loop level through kinetic mixing with the photon
and the $Z^0$.  These signals have been searched for by a number
of gamma-ray telescopes, and limits have been placed on the corresponding
gamma-ray fluxes.  We examine here the constraints on the $\chi_1^x$ 
state from observations of the galactic centre~(GC) gamma ray continuum by the Fermi Large
Area Telescope~(Fermi-LAT)~\cite{fermigc}, 
as well as from observations of the diffuse
photon flux by the INTEGRAL~\cite{Bouchet:2008rp},
COMPTEL~\cite{COMPTEL}, EGRET~\cite{EGRET}, and Fermi~\cite{FermiDiffuse}
experiments.  For the GeV-scale dark matter masses we are considering,
these observations are expected to give the strongest 
constraints~\cite{Hooper:2012sr,Essig:2013goa}.
\footnote{We have also examined constraints from monochromatic photon   
line searches and found the continuum constraints significantly more stringent for the small values of $\epsilon$ allowed by fixed target experiments.}  
We also study bounds
from the effects of DM annihilation during recombination on the cosmic
microwave background~(CMB)~\cite{Padmanabhan:2005es,Slatyer:2009yq}.

  The continuum photon flux from $\chi_1^x$ pair
annihilation into hidden vectors is given by 
\begin{equation} \label{eq:flux}
\frac{d\Phi_{\gamma}}{d E_{\gamma}}= \frac{\langle \sigma v\rangle_{\chi \chi
\rightarrow XX}}{8 \pi m_{\chi}^2}  \frac{d N^{\rm tot}_{\gamma}}{d E_{\gamma}}
\times \int\!dl\;\rho^2(l)  \ , 
\end{equation} 
where $\langle\sigma v\rangle_{\chi\chi\to XX}$ is the thermally 
averaged annihilation rate at present, $\rho(l)$ is the dark matter
density along the line of sight $l$, and $dN^{\rm tot}_{\gamma}/dE_{\gamma}$ is
the total differential photon yield per annihilation, defined as
\begin{equation} 
\frac{d N^{\rm tot}_{\gamma}}{d E_{\gamma}} \equiv \sum_f \BR_f
\frac{d N^f_{\gamma}}{d E_{\gamma}} 
\end{equation} 
where $\BR_f$ is the branching
fraction of the $XX$ state into the final state $f$.

  In our calculations, we use the results of 
Refs.~\cite{Meade:2009rb, Hooper:2012cw} 
to estimate the partial yields $dN^f_{\gamma}/dE_{\gamma}$
by interpolating between the results for the values of
$m_{\chi}$ and $m_{\chi}/m_x$ listed in these studies. 
For the dark matter density profile, we consider four distributions
that span the range of reasonable possibilities:
Navarro-Frenk-White~(NFW)~\cite{Navarro:1995iw}, Einasto~\cite{Navarro:2003ew,
Springel:2008cc}, contracted~\cite{Hooper:2012sr}, 
and cored NFW~\cite{Hooper:2012sr}.  These take the forms
\beq \label{eq:profiles}
\rho(r) \propto \left\{
\begin{array}{lcl}
\left[\frac{r}{R_s}\left(1+\frac{r}{R_s}\right)^2\right]^{-1}&~~&\text{(NFW)}\\
e^{-2/\alpha\left[\left(\frac{r}{R_s}\right)^\alpha -1\right]}&~~&\text{(Einasto)}\\
\left[\left(\frac{r}{R_s}\right)^{\gamma}
\left(1+\frac{r}{R_s}\right)^{3-\gamma}\right]^{-1}&~~&\text{(contracted)}\\
\left[\frac{r_c+(r-r_c)
\hspace{.1cm}\Theta(r-r_c)}{R_s}\left(1+\frac{r_c+(r-r_c) \hspace{.1cm}
\Theta(r-r_c)}{R_s}\right)^2\right]^{-1}&~~&\text{(cored)}
\end{array}
\right. \ .
\eeq
Here, $r$ is the radial distance from the GC and $\Theta$ is a step function.  
Following Refs.~\cite{Fan:2013faa,Hooper:2012sr}, we fix the scale
radius to be $R_s=20$\,kpc and the Einasto parameter $\alpha=0.17$.
For the contracted profile we set $\gamma=1.4$ and for the cored profile
we set the core radius to be $r_c=1$\,kpc, as in Ref.~\cite{Hooper:2012sr}.  
In all four cases, we fix the overall normalization such that 
$\rho(r=8.5\,\text{kpc})=0.3\,\gev/\text{cm}^3$.

  Using these halo profiles, we are able to compute the gamma-ray fluxes
from hidden dark matter created in moduli decays and compare them
to limits derived from observations of the GC and the diffuse gamma-ray
background.  
For the GC signal, we use the limits on $\langle \sigma v \rangle/m_{\chi}^2
\int_{E_{\rm min}}^{E_{\rm max}}\!\!dE_{\gamma}\;dN_{\gamma}^{\rm tot}/dE_{\gamma}$ 
computed in Ref.~\cite{Hooper:2012sr} in several energy bins 
$[E_{{\rm min},i},E_{{\rm max},i}]$ and each of the four
DM profiles described above.   For the diffuse gamma ray background,
we use the flux limits compiled and computed in Ref.~\cite{Essig:2013goa}.

  In addition to measurements of cosmic gamma rays, observations of the 
CMB also provide a significant limit on 
DM annihilation~\cite{Padmanabhan:2005es,Slatyer:2009yq}.
The energy released by dark matter annihilation around the time of recombination
will distort the last scattering surface, and hence affect the CMB anisotropies.
The limit derived from this effect 
is~\cite{Hutsi:2011vx,Galli:2011rz,Finkbeiner:2011dx}
\beq
f \frac{\Omega^2_{\chi}}{\Omega^2_{\rm cdm}}\langle\sigma v\rangle_{\rm CMB} 
\leq (2.42\times 10^{-27}\,{\rm cm}^3/{\rm s})\lrf{m_{\chi}}{\gev} \ ,
\label{eq:cmb_bound_sym}
\eeq
where ${\langle\sigma v\rangle_{\rm CMB}}$ is the thermally averaged
cross section during recombination
and $f$ is a constant efficiency factor parametrizing the fraction of energy 
transferred to the photon-baryon fluid, which can typically range 
from $f \approx 0.2\!-\!1.0$~\cite{Finkbeiner:2011dx}. We will vary  
$f$ across this range to illustrate its effect on the resulting constraint.

\begin{figure}[ttt]
\centering
\includegraphics[width=8cm]{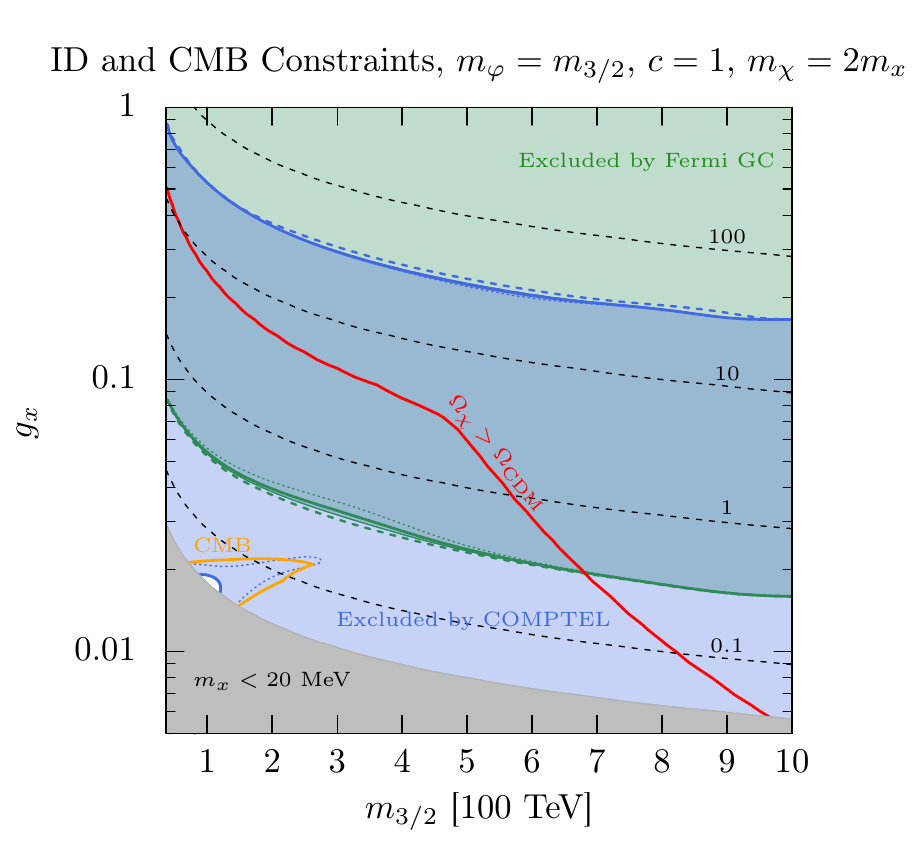}
\includegraphics[width=8cm]{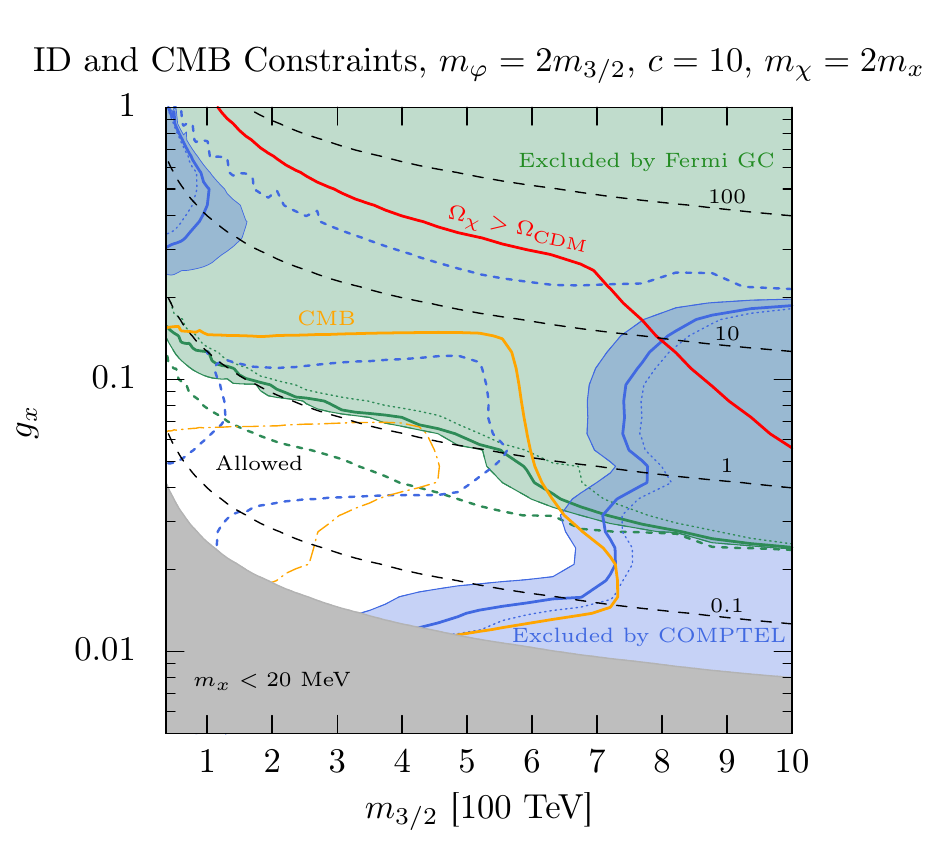}
\caption{Constraints from indirect detection on hidden $U(1)_x$
neutralino DM produced by moduli decays for $m_x=m_{\chi}/2$,
$\xi = 0.1$, as well as $(c\!=\!1,\,\mphi\!=\!\mgrav)$~(left), and 
$(c\!=\!10,\,\mphi\!=\!2\mgrav)$~(right).
The green shaded region is excluded by Fermi GC observations
and the blue shaded region is excluded by COMPTEL.  
Both exclusions assume an Einasto galactic DM profile.
The thick solid and thin dotted contours correspond to the exclusions 
assuming the NFW and cored profiles, respectively.
The green and blue dashed lines show the boundaries of
the stronger exclusion obtained assuming a contracted profile with 
$\gamma = 1.4$.  Above and to the right of the solid red line, 
the hidden LSP density is larger than the observed DM density.  
The solid and dash-dotted orange lines shows the exclusion from deviations 
in the CMB for $f=0.2$ and $f=1$, respectively, with the excluded areas 
above and to the right of the lines.  
Note that the entire $c=1$ parameter space is
excluded by the CMB constraint for $f=1$.
The gray shaded region at the bottom has a hidden vector mass 
$m_x < 20\,\mev$ that is excluded by fixed-target experiments.}
\label{fig:u1symid}
\end{figure}

  These observations put very strong constraints on 
hidden dark matter when it is produced in moduli decays.
The corresponding ID and CMB bounds are shown in Fig.~\ref{fig:u1symid}
in the $m_{3/2}-g_x$ plane.  We fix the 
moduli parameters to $\mphi=\mgrav$ and $c=1$ in the left panel
and $\mphi=2\mgrav$ and $c=10$ in the right.  The relevant hidden-sector
parameters are taken to be $\xi = 0.1$ and $m_x=m_{\chi}/2$.
The solid red line shows where $\Omega_{\chi} = \Omega_{\rm cdm}$,
with the region above and to the right of the line producing too
much dark matter.  The green shaded regions show the exclusion 
from Fermi observations of the GC assuming the Einasto DM profile 
of Eq.~\ref{eq:profiles} rescaled by the expected dark matter fraction 
$(\Omega_{\chi}/\Omega_{\rm cdm})^2$, while the blue shaded regions 
show the exclusion from COMPTEL under the same conditions.  
Exclusions for other profiles are also shown by the parallel
contours.\footnote{The thick green and blue dashed lines show the
boundaries of the regions excluded for a more aggressive contracted profile
with $\gamma = 1.4$.  For clarity, we do not shade the interior 
of these. The thick solid and thin dotted contours correspond 
to the NFW and cored profiles, respectively.}  
We have also considered the corresponding constraints from INTEGRAL, EGRET, 
and Fermi diffuse gamma ray observations, but these do not exclude 
any additional parameter space and so are not included 
in Fig.~\ref{fig:u1symid} for clarity.
Limits from CMB distortions are shown by the solid and dash-dotted 
orange lines, for $f=0.2$ and $1$ respectively, with the 
excluded region above and to the right of the contours.
The dashed black lines are contours of the hidden LSP $\chi$ mass in GeV,
with the region where $m_x = m_{\chi}/2 < 20\,\mev$ excluded by fixed
target experiments~\cite{Bjorken:2009mm}.

 For generic moduli parameters, $c=1$ and $\mphi = \mgrav$,
we find that constraints from indirect detection and CMB observations 
nearly completely rule out this scenario even with optimistic choices 
for the DM halo properties and CMB energy injection efficiency.  
However, for 
$c=10$ and $\mphi = 2\mgrav$, the hidden neutralino relic density can
become sufficiently small to evade the strong limits from ID and the CMB,
despite the relatively large $\chi_1^{x}$ annihilation cross section.
In this case, a second more abundant contribution to the total dark
matter abundance would be needed.  Note as well that the remaining
allowed region corresponds to sub-GeV hidden sector masses that
could potentially be probed in current and planned precision
searches~\cite{Essig:2013lka}.

\subsection{Summary}

    With optimistic but reasonable choices for the moduli parameters,
a light hidden sector neutralino LSP produced in moduli reheating
can be consistent with current DM searches.  Even so, the
scenario is tightly constrained by indirect detection 
and CMB measurements.   The challenge here is precisely the same as 
in the MSSM: to avoid overproducing the neutralino LSP during
moduli reheating, the annihilation rate must be large relative
to the standard thermal value
$\langle \sigma v\rangle \sim 3\times 10^{-26}\,\text{cm}^3/\text{s}$,
and such an enhanced rate is strongly constrained by indirect DM searches.  
To avoid these bounds while not creating too much dark matter, the annihilation
rate must be large enough that the LSP relic abundance is only a small fraction
of the total DM density.

  The $U(1)_x$ hidden sector does slightly better than the MSSM in this regard
for two reasons.  First, the hidden gauge coupling can be taken small
(as can $\xi = \mu'/M_x$), which helps to reduce the LSP relic abundance
as suggested by Eq.~\eqref{eq:omscale}.  And second, the hidden LSP can
be much lighter than an MSSM wino or Higgsino, leading to smaller photon
yields below the primary sensitivity of Fermi-LAT.  The strongest 
constraints for such light masses come from COMPTEL, which are less stringent
than those from Fermi.  Since the large late-time hidden neutralino 
annihilation rate is the primary hindrance to realizing this set-up, 
one might consider  analogous scenarios in which the CMB and indirect detection 
signatures are suppressed; we address this possibility in the following section.

  Before moving on, let us also comment on the spectrum in the hidden sector.  
To avoid a large fine tuning, the hidden scalar soft terms must be 
relatively small, on the same order or less than the hidden gaugino mass.
Given the large values of $\mgrav$ considered, the scalar soft masses
must be sequestered from supersymmetry breaking.  They must also receive
new contributions beyond minimal AMSB, and the $b'$ bilinear soft term
must not be too much larger than $(\mu')^2$.  All three features require 
non-trivial additional structure in the underlying mechanisms 
of supersymmetry breaking or mediation~\cite{Pomarol:1999ie,Katz:1999uw}.

\section{Variation \#2: Asymmetric Hidden $U(1)$\label{sec:u1asym}}

  As a second extension of the MSSM, we investigate a theory
of hidden asymmetric dark matter~(ADM)~\cite{
Nussinov:1985xr,Barr:1991qn,Kaplan:1991ah,Kaplan:2009ag}. 
In the ADM framework, the DM particle has a distinct antiparticle,
and its abundance is set mainly by a particle-antiparticle 
asymmetry in analogy to baryons, and this tends to suppress indirect
detection signals from late-time annihilation if very little
anti-DM is present~\cite{Graesser:2011wi,Iminniyaz:2011yp,Lin:2011gj, Bell:2014xta}.
The ADM theory we consider is nearly identical to the hidden
$U(1)_x$ theory studied in Section~\ref{sec:u1sym}, but with
an additional pair of vector-like hidden chiral 
superfields $Y$ and $Y^c$ with $U(1)_x$ charges $x_Y=\pm 1$.  
We assume that a small asymmetry in the $Y$ density is generated
during moduli reheating, in addition to the much larger symmetric density, 
and we compute the resulting relic densities and experimental signals.

\subsection{Mass Spectrum and Decays\label{sec:u1asym_massspec}}

  The superpotential in the hidden sector is same as that 
considered in Sec.~\ref{sec:u1sym} up to a new mass 
term for the $Y$ and $Y^c$ multiplets,
\beq
W \supset -\mu_Y YY^c \ .
\eeq
We also include the new soft supersymmetry breaking terms
\beq
-\mathscr{L}_{\mathrm{soft}} \supset m_{\yt}^2|\yt|^2+m_{\ytc}^2|\ytc|^2
- (b_Y\widetilde{Y}\widetilde{Y}^c + \text{h.c.}) \ .
\eeq
As in Sec.~\ref{sec:u1sym}, we fix the hidden gaugino mass
to its AMSB value with $b_x = 2(1+1)$, accounting for the new superfields. 
We also do not impose minimal AMSB values for the scalar soft terms,
but take them (as well as $\mu'$ and $\mu_Y$) to be of similar size to the 
gaugino soft mass.  Finally, we arrange parameters so that
the hidden Higgs scalars develop expectation values and spontaneously
break the $U(1)_x$.

  The mass spectrum of the hidden sector follows the
minimal model considered in Sec.~\ref{sec:u1sym}, but now a 
new Dirac fermion $\Psi$ of mass $m_{\Psi}=\mu_Y$ and 
two complex scalars $\Phi_{1,2}$.  
The scalar mass matrix in the $(\widetilde{Y},\,\widetilde{Y}^{c*})$ 
basis is
\beq
\mathcal{M}^2_{\widetilde{Y}} = \left(
\begin{array}{cc} |\mu_Y|^2 + m^2_{\yt} - \tilde{\delta}_D & b_Y^* \\ 
b_Y & |\mu_Y|^2 + m_{\ytc}^2 + \tilde{\delta}_D,
\end{array}
\right) \ ,
\label{eq:scalarmassmat}
\eeq
where $\tilde\delta_D = g_x^2 \eta^2 \cos 2\zeta + x_Y \epsilon g_x g' v^2 \cos 2\beta/2$.
Taking $m_{\yt}^2 = m_{\ytc}^2$ for convenience, the mass eigenvalues are
\beq
m^2_{1,2} = |\mu_Y|^2 + m^2_{\yt} \mp \sqrt{\tilde\delta_D^2 + |b_Y|^2} \ .
\label{eq:Yeigenmass}
\eeq
In what follows we will refer to the lighter scalar $\Phi_1$ as $\Phi$.

  This theory preserves both the usual $R$-parity as well as
a non-anomalous global $U(1)$ flavour symmetry 
among the $Y$ and $Y^c$ multiplets, and can support multiple stable states.
The number of stable particles depends on the mass spectrum.
To allow for dominantly asymmetric dark matter, we will focus on
spectra with $m_{\chi_1^x} > m_{\Psi}+m_{\Phi}$ such that the decay
$\chi_1^x\to \Psi +\Phi^*$ is possible, and the only stable
hidden states are $\Psi$ and $\Phi$.
If this channel is not kinematically allowed, the $\chi_1^x$ neutralino 
will also be stable and can induce overly large gamma ray signals as in 
the previous section.  
We also choose soft masses such that $m_x < m_\Phi,\,m_{\Psi}$ 
to allow both states to annihilate efficiently into hidden vectors.
With this mass ordering, the lightest hidden states will
be the vector $X^{\mu}$ and the hidden Higgs $h_1^x$.  
Both will decay to the SM in the same way as in 
the minimal model of Sec.~\ref{sec:u1sym}.
The lightest MSSM neutralino will also continue to decay to 
the hidden sector through gauge kinetic mixing, now with
additional decay modes $\chi_1^0\to \Psi\Phi_{1,2}$.  
As before, the net $\chi_1^0$ lifetime is expected to be
short relative to the cosmological timescales of interest.

\subsection{Moduli Reheating and Asymmetric Dark Matter}

  The $\Psi$ and $\Phi$ states will both act as ADM if they
are created in the moduli reheating process slightly more often
than their antiparticles.  
The production of the asymmetry can be accommodated within a set of
Boltzmann equations similar to Eq.~\eqref{eq:boltzdm}
as follows:
\beq
\frac{dn_{\Psi}}{dt} + 3Hn_{\Psi}&=& 
(1+\kappa/2)\frac{\mathcal{N}_{\Psi}\Gamma_{\varphi}}{\mphi}\rho_{\varphi}
-\langle\sigma v\rangle_{\Psi}\left(n_{\Psi}n_{\overline{\Psi}}-(n_\Psi^{\mathrm{eq}})^2\right) \\ 
 & - &  \langle \sigma v \rangle_\mathrm{trans}
(n_\Psi^2 -\nu^2 n_{\Phi}^2) \notag\\
\frac{dn_{\Phi}}{dt} + 3Hn_{\Phi}&=& 
(1+\kappa/2)\frac{\mathcal{N}_{\Phi}\Gamma_{\varphi}}{\mphi}\rho_{\varphi}
-\langle\sigma v\rangle_{\Phi}\left(n_{\Phi}n_{\Phi^*}-(n_\Phi^{\mathrm{eq}})^2\right)\\ 
 &-&  \langle \sigma v \rangle_\mathrm{trans}
(\nu^2 n_\Phi^2 -n_{\Psi}^2) \notag,
\eeq    
with a similar set of equations for the anti-DM $\overline \Psi$ and $\Phi^*$, 
but with $\kappa \rightarrow -\kappa$.  
Here, $\mathcal{N}_{\Psi}$ and $\mathcal{N}_{\Phi}$ are the mean number 
of $\Psi$ and $\Phi$ produced per modulus decay.  
This includes particles created directly in moduli decays, rescattering, 
and from the cascade decays of other states.  
The thermally-averaged cross sections $\langle\sigma v\rangle_{\Psi,\Phi}$
describe the $\Psi\overline{\Psi}$ and $\Phi\Phi^*$ annihilation,
while $\langle\sigma v\rangle_{\mathrm{trans}}$ in each equation corresponds 
to the transfer reaction 
$\Psi\Psi\leftrightarrow\Phi\Phi$ mediated by $U(1)_x$ gaugino exchange
with $\nu = 2\left(m_\Psi/m_\Phi\right)^2 K_2(m_\Psi/T)/K_2(m_{\Phi}/T)$.

  Asymmetry generation in this scenario is parametrized by the
constant $\kappa$.  It could arise directly from moduli decays
or from the interactions of intermediate moduli decay products
along the lines of one of the mechanisms of 
Refs.~\cite{Davoudiasl:2010am,Allahverdi:2010rh,
Bell:2011tn,Cheung:2011if,Kane:2011ih,Zurek:2013wia,Fischler:2014jda}.
Indeed, this theory can be viewed as a simplified realization of
the supersymmetric hylogenesis model studied in Ref.~\cite{Blinov:2012hq}.
Relative to that work, we undertake a more detailed investigation
of the relic density resulting from different choices for the 
moduli parameters, and we do not attempt to link the DM asymmetry 
to the baryon asymmetry.

  The annihilation cross section $\langle\sigma v\rangle_{\Psi}$ 
is dominated by the ${\Psi}\overline{\Psi}\to XX$ channel to hidden vector
bosons and is given by
\begin{align}
\langle\sigma v\rangle_\Psi & ~=~   
\frac{1}{16\pi}\frac{g_x^4}{m_\Psi^2}
\left(1-\frac{m_x^2}{m_\Psi^2}\right)^{3/2}\left(1-\frac{m_x^2}{2m_\Psi^2}\right)^{-2}\\
& ~\simeq~  (1.5\times 10^{-24}\,\text{cm}^2/\text{s})
\left(\frac{g_x}{0.05}\right)^4 
\left(\frac{1\,\GeV}{m_\Psi}\right)^2.
\end{align} 
The scalar annihilation rate is similar. 
For the transfer reaction, we have
\beq
\langle\sigma v\rangle_\text{trans} \approx  \frac{g_x^4}{8\pi}
\sqrt{1-\frac{m_\Phi^2}{m_\Psi^2}}
\left|
\sum_{k=1}^3 \frac{(A_k^{*2} - B_k^2) m_{\chi^x_k}}
{m_{\chi^x_k}^2 + m_\Psi^2 - m_\Phi^2}
\right|^2\ ,
\eeq
where $A_k = \mathbf{Z}_{11}^* \mathbf{P}_{k3}$ and 
$B_k = \mathbf{Z}_{12} \mathbf{P}_{k3}$ with  
$\mathbf{P}_{k3}$ the HS gaugino content of $\chi^x_k$
and $\mathbf{Z}_{ij}$ is a unitary matrix that diagonalizes
the scalar mass matrix of Eq.~\eqref{eq:scalarmassmat}.  
Note that the transfer reaction can be suppressed relative 
to annihilation for $m_{\chi_1^x} > m_{\Psi}+m_{\Phi}$.

\subsection{Relic Densities and Constraints}

  To investigate the relic densities of $\Psi$ and $\Phi$ in this
theory following moduli reheating and the corresponding constraints
upon them, we set all the dimensionful hidden parameters to be fixed ratios 
of the $U(1)_x$ gaugino soft mass 
$M_x = 4g_x^2m_{3/2}/(4\pi)^2$:
\beq
m_{A^x} = 10 \mu' = 50 \mu_Y = 100 m_x = 250 b_Y^{1/2} = 250 m_{\widetilde{Y}} = M_x 
\ .
\eeq
With these choices, the mass spectrum for $g_x=0.1$ and $\mgrav=200\,\tev$ is
\beq
m_{\Psi} =1\gev,~
m_{\Phi} =0.97\gev,~
m_{\chi_1^x} =5.1\gev,~
m_x = 0.51\gev,~
m_{h_1^x} = 0.5\gev \ .
\nnmb
\eeq
This mass ordering coincides with the spectrum described in
in Sec.~\ref{sec:u1asym_massspec}.  

\begin{figure}[ttt]
\centering
\includegraphics[width=10cm]{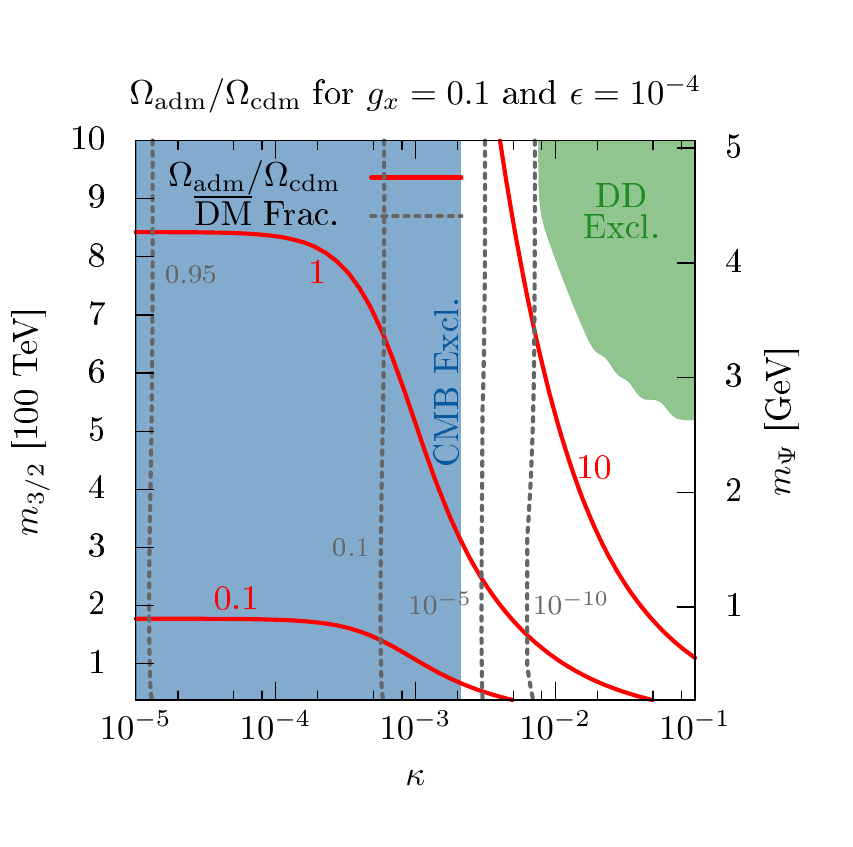}
\caption{Abundance of $\Psi$ and $\Phi$ in the $\kappa - m_{3/2}$ plane. 
The right $y$ axis shows the $\Psi$ mass $m_\Psi = \mu_Y$. 
Solid red contours show the fraction of the measured abundance made up 
by $\Psi$ and $\Phi$ and their anti-particles. The dashed grey 
lines show the fractional asymmetry between DM and anti-DM. 
The blue region is excluded by the CMB bound and the green by direct detection. 
\label{fig:adm_scan1}}
\end{figure}

  In Fig.~\ref{fig:adm_scan1} we show the dark matter abundance 
$\Omega_{\rm adm} = \rho_\mathrm{adm}/\rho_c$ of $\Psi$ and $\Phi$ 
(and their antiparticles)
relative to the observed abundance $\Omega_\mathrm{cdm}$ in the 
$\kappa\!-\!\mgrav$ plane for $g_x=0.1$, $\epsilon=10^{-4}$,
$\mphi=\mgrav$, and $c=1$. 
Contours of $\Omega_\mathrm{adm}/\Omega_\mathrm{cdm} = 0.1,\,1,\,10$ are given
by solid red lines.  The grey dashed lines in this figure
correspond to the net residual anti-DM abundance $R_{\Phi}+R_{\Phi}$, 
where $R_{\Psi} = \Omega_{\overline{\Psi}}/\Omega_{\Psi}$
and similarly for $\Phi$.  Not surprisingly, larger values of the production
asymmetry parameter $\kappa$ lead to smaller residual anti-DM abundances.
In this figure we also show in blue the region of parameters that is excluded by
CMB observations, as well as the region excluded by direct detection in green.
These constraints will be discussed in more detail below.

\begin{figure}[ttt]
\centering
\includegraphics[width=10cm]{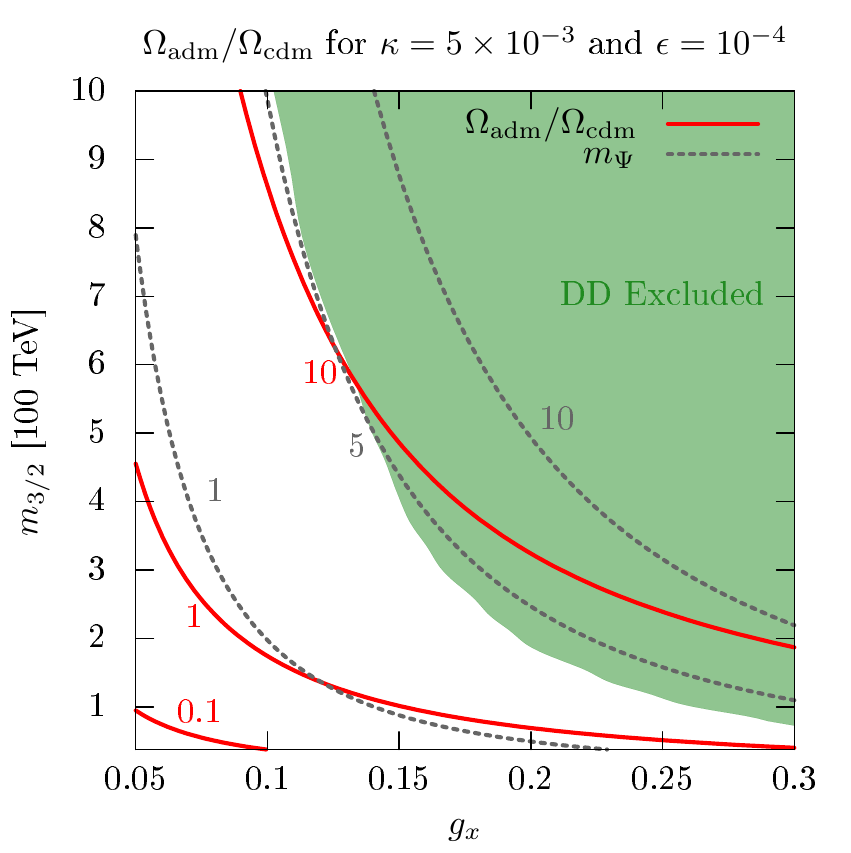}
\caption{Abundance of $\Psi$ and $\Phi$ in the $g_x - m_{3/2}$ plane. 
Solid red contours show the fraction of the measured abundance made up 
by $\Psi$ and $\Phi$ and their anti-particles. The dashed grey 
lines show the $\Psi$ mass in GeV. 
The green region is excluded by direct detection. 
\label{fig:adm_scan2}}
\end{figure}

  The ADM abundance in the $g_x\!-\!\mgrav$ plane is
shown in Fig.~\ref{fig:adm_scan2} for $\kappa = 5\times 10^{-3}$,
$\epsilon=10^{-4}$, $\mphi=\mgrav$, and $c=1$. 
Again, contours of $\Omega_\mathrm{adm}/\Omega_\mathrm{cdm} = 0.1,\,1,\,10$ are
given by solid red lines.  We also plot contours of the $\Psi$ mass
with dashed grey lines.  As before, the shaded green region
is excluded by direct detection searches.

  The region excluded by CMB observations in Fig.~\ref{fig:adm_scan1}
(shaded blue) coincides with larger values of the residual anti-DM
abundances $R_{\Psi}+R_{\Phi}$.  These residual abundances provide
an annihilation mode that injects energy into the
cosmological plasma during the CMB era~\cite{Lin:2011gj},
as discussed in Sec.~\ref{subsec:id}.  Accounting for exclusively asymmetric
annihilation and the multiple DM species, the result
of Eq.~\eqref{eq:cmb_bound_sym} translates into
\beq
2f\sum_{i=\Psi,\Phi}\left(\frac{\Omega_{i}+\Omega_{\bar i}}{\Omega_\mathrm{cdm}}\right)^2 
\frac{R_i}{(1+R_i)^2}\,\frac{\langle \sigma v \rangle_{i}}{m_i} ~~<~~
\frac{2.42\times 10^{-27}\;\mathrm{cm^3/s}}{\GeV}.
\label{eq:cmb_bound}
\eeq
The CMB exclusion shown in Fig.~\ref{fig:adm_scan1} uses $f=1$, but other
values in the range $f=0.2\!-1.0$ yield similar results.  The boundary of 
the excluded region is also nearly vertical and independent of $\mgrav$.
This can be understood in terms of an approximate cancellation of factors
of $\mgrav = \mphi$ in the combination 
$\Omega_{\rm adm}^2\langle\sigma v\rangle/m$,
while $R_i$ is determined primarily by $\kappa$.
In addition to the limits from the CMB, we have also computed the bounds from
indirect detection as described in Sec.~\ref{sec:u1sym}.  
These searches yield exclusions very similar to that from the CMB 
and are omitted from Fig.~\ref{fig:adm_scan1}.

  Direct detection searches also place a significant constraint on this
ADM scenario.  Kinetic mixing of the hidden $U(1)_x$ with hypercharge
links the hidden vector to charged matter with an effective
coupling proportional to $-e\epsilon\,c_W$.  In the present case,
the dark matter consists of Dirac fermions and complex scalars charged
under $U(1)_x$, and this allows a vectorial coupling of these states
to the $X$ gauge boson.  Together, these two features induce
a vector-vector effective operator (for $m_x \gtrsim 20\,\mev$) 
connecting the DM states to the proton
that gives rise to spin-independent~(SI) scattering on nuclei.
The $\Psi$-proton scattering cross section is
\beq
\sigma_p = \frac{\epsilon^2 c_W^2 e^2 g_x^2 \mu_n^2}{\pi m_x^4} \ .
\label{eq:protxsec}
\eeq 
A similar expression applies to the scalar $\Phi$.  
This gives rise to an effective SI cross section per nucleon
(in terms of which experimental limits are typically quoted) of
\beq
\tilde \sigma_n &=& (Z^2/A^2) \sigma_p 
\label{eq:dd_xsec}\\
&\simeq& 2\times 10^{-38}\;\mathrm{cm^2} 
\left(\frac{2Z}{A}\right)^2 
\left(\frac{\epsilon}{10^{-3}}\right)^2 
\left(\frac{g_x}{0.1}\right)^2 
\left(\frac{\mu_n}{1\;\GeV}\right)^2 
\left(\frac{1\;\GeV}{m_x}\right)^4 \ .
\nnmb
\eeq
Comparing this result to the exclusions of low-mass DM from
LUX~\cite{Akerib:2013tjd}, XENON10 S2 only analysis~\cite{Angle:2011th}, 
CDMSLite~\cite{Agnese:2013jaa} and CRESST-Si~\cite{Angloher:2002in},
we obtain the green exclusion regions shown in Figures~\ref{fig:adm_scan1} and
\ref{fig:adm_scan2}.

\subsection{Summary}

  This hidden $U(1)_x$ extension of the MSSM can account for the entire
relic dark matter abundance in the aftermath of moduli reheating
while being consistent with existing constraints from direct and indirect
detection.  Even though the DM annihilation cross section is much
larger than the standard thermal value, a strong DM-anti-DM asymmetry
allows for a significant total density while suppressing DM annihilation
signals at late times.  
Limits from direct detection searches can also
be evaded for light DM masses below the sensitivity of current
experiments.   

  To achieve a strong DM asymmetry, a relatively large asymmetry parameter
$\kappa \gtrsim 10^{-3}$ is needed.  We have not specified the dynamics
that gives rise to the asymmetry in moduli reheating, but more complete
theories of asymmetry generation suggest that values this large 
can be challenging to obtain~\cite{Allahverdi:2010rh,Bell:2011tn,
Cheung:2011if,Blinov:2012hq}.  Furthermore, as in the symmetric
hidden sector theory considered previously, the spectrum required for this 
mechanism to work requires scalar sequestering and scalar soft masses
of the right size.

\section{Variation \#3: Hidden $SU(N)$\label{sec:sun}}

  The third extension of the MSSM that we consider consists of a pure
supersymmetric $SU(N)_x$ gauge theory together with heavy 
connector matter multiplets charged under both $SU(N)_x$  
and the MSSM gauge groups.\footnote{See also Refs.~\cite{Feng:2011ik,Boddy:2014yra} 
for previous studies of this scenario in a slightly different context.}
In contrast to the two previous extensions, we do not have to make 
any strong assumptions about the scalar soft mass parameters  
for the theory to produce an acceptable LSP relic density.  
In particular, this extension can work in the context 
of a \emph{mini-split} 
spectrum where the scalar superpartners are much heavier 
than the gauginos~\cite{Wells:2004di,Hall:2011jd,Arvanitaki:2012ps,ArkaniHamed:2012gw,Ibe:2011aa,Ibe:2012hu}.

\subsection{$SU(N)_x$ Mass Spectrum and Confinement}

  The hidden states below the TeV scale consist of the 
$SU(N)_x$ gluon and gluino.
The hidden gluino soft mass is
\beq
M_x = {r_x}\frac{g_x^2}{(4\pi)^2}m_{3/2} \ ,
\label{mgluino}
\eeq
where $r_x = 3N$ if it is generated mainly by AMSB effects.
In the discussion to follow, we will consider additional heavy matter
charged under $SU(N)_x$ with large supersymmetric mass $\mu_F$.
For $\mu_F \gg m_{3/2}$, the coefficient $r_x$ will be 
unchanged~\cite{Giudice:1998xp}.
However, when $\mu_F \lesssim m_{3/2}$, the value of $r_x$ can be modified 
by an amount of order unity that depends on the soft masses of these 
states~\cite{Sundrum:2004un,Gupta:2012gu}.  
We consider deviations in $r_{x}$ away
from the AMSB value but still of the same general size.

    Below the hidden gluino mass, the hidden sector is a pure
$SU(N)_x$ gauge theory.  It is therefore guaranteed to be asymptotically
free, and the low-energy theory of hidden gluons should undergo a
confining transition at some energy scale $\Lambda_x$ to a theory
of massive glueball (and glueballino) bound states.  
The one-loop estimate of the confinement scale gives
\beq
\Lambda_x = M_x\exp\left(-\frac{3r_x}{22N}\frac{m_{3/2}}{M_x}\right) \ .
\eeq
Demanding that the $SU(N)_x$ gluino be lighter than the lightest
MSSM neutralino typically forces $\Lambda_x$ to be very small.
For example, setting $M_x < 1000\,\gev$, $r_x=3N$, and requiring
that $M_x < M_2$ (with its value as in AMSB, $M_2 \simeq m_{3/2}/360$),
one obtains $\Lambda_x < 10^{-61}\,\gev$.  Thus, we will neglect
$SU(N)_x$ confinement in our analysis and treat the hidden gauge theory
as weakly interacting.

\subsection{Connectors to the MSSM}

  The lightest MSSM superpartner must be able to decay to the hidden
sector for this extension to solve the MSSM moduli relic problem.
Such decays can be induced by heavy matter multiplets charged under 
both the MSSM gauge groups and $SU(N)_x$.  Following Ref.~\cite{Feng:2011ik},
we examine two type of connectors.

  The first set of connectors consists of $N_F$ pairs 
of chiral superfields $F$ and $F^c$ with charges $(1,2,\mp1/2;N)$ 
under $SU(3)_c\!\times\!SU(2)_L\!\times\!U(1)_Y\!\times\!SU(N)_x$
with a supersymmetric mass term~\cite{Feng:2011ik}
\beq
W \supset \mu_FFF^c \ .
\eeq
For $\mu_F \gtrsim m_{3/2}$, the heavy multiplets can be integrated out
supersymmetrically to give~\cite{Feng:2011ik} 
\beq
-\Delta\mathscr{L} &\supset& 
\int\!d^4\theta\;\frac{g_x^2g_2^2}{(4\pi)^2}\frac{2N_F}{\mu_F^4}\,
{{W}^{\dagger}_{x\,\dot{\alpha}}}{W}^{\dagger\,\dot{\alpha}}
{W}^{{\alpha}}_x{W}_{{\alpha}}\\
&\supset& \alpha_x\alpha_2\frac{2N_F}{\mu_F^4}
\left[{\widetilde{G}}_x^{\dagger}(\bar{\sigma}\ccdot\del)\widetilde{W}\;
G^{\mu\nu}_xW_{\mu\nu} 
+ \left(G^{\mu\nu}_xW_{\mu\nu}\right)^2 
\right] \ .
\label{wop1}
\eeq
Similar operators involving the $U(1)_Y$ vector multiplet will also
be generated, and additional operators will also arise
with the inclusion of supersymmetry breaking.
The wino operator of Eq.~\eqref{wop1} allows the decay 
$\widetilde{W}^0 \to W^0\,G_x\,\widetilde{G}_x$, whose rate we estimate to be
\beq
\Gamma &\sim& \frac{4(N^2\!-\!1)N_F^2}{8\pi(4\pi)^2}\,\alpha_x^2\alpha_2^2\,
|\mathbf{N}_{12}|^2\frac{m_{\chi_1^0}^9}{\mu_F^8} \\
&\simeq& (7\times 10^{5}\,\text{s})^{-1}(N^2\!-\!1)N_F^2|\mathbf{N}_{12}|^2
\lrf{\alpha_x}{10^{-3}}^2
\lrf{m_{\chi_1^0}}{270\,\gev}^9
\lrf{100\,\tev}{\mu_F}^8 \ ,
\nnmb
\eeq
where $m_{\chi_1^0}$ is the mass of the lightest MSSM neutralino,
$|\mathbf{N}_{12}|$ is its wino content, and the fiducial value of 
$m_{\chi_1^0}$ corresponds to the AMSB value of $M_2$ for 
$\mgrav\simeq 100\,\tev$.
Note that these sample parameter values lead to decays after
the onset of primordial nucleosynthesis. 

  The second set of connectors that we consider consists of the same 
$N_F$ heavy multiplets $F$ and $F^c$ together with $P$ and $P^c$ multiplets 
with charges $(1,1,0,\bar{N})$~\cite{Feng:2011ik}.  
This allows the couplings
\beq
W \supset 
\lambda_uH_uFP 
+ \lambda_dH_dF^cP^c 
+ \mu_FFF^c+\mu_PPP^c \ .
\eeq
Neglecting supersymmetry breaking, integrating out the heavy $F$ and $P$
multiplets at one-loop order generates operators such as~\cite{Feng:2011ik}
\beq
-\Delta\mathscr{L} &\supset& 
\int\!d^2\theta\;\frac{g_x^2\lambda_u^2}{(4\pi)^2}\frac{2N_F}{\mu_F^2}\,
{W}^{x\,{\alpha}}{W}_{x\,\dot{\alpha}}\;H_u\ccdot H_d\\
&\supset& \alpha_x\lrf{\lambda_u^2}{4\pi}\frac{2N_F}{\mu_F^2}
\left[{\widetilde{G}}_x\sigma_{\mu}\bar{\sigma}_{\nu}\widetilde{H}_d\,
H_u\,G_x^{\mu\nu} + G_x^{\mu\nu}G_{x\,\mu\nu}\,H_u\ccdot H_d\right] \ .
\label{wop2}
\eeq
where we have set $\mu_P=\mu_F$ and $\lambda_d=\lambda_u$ for simplicity.
Additional related operators arise when supersymmetry breaking is included.
The first term in Eq.~\eqref{wop2} induces the decay 
$\chi_1^0\to G_x\tilde{G}_x$, whose rate we estimate to be
\beq
\Gamma &\sim& \frac{4(N^2\!-\!1)N_F^2}{8\pi}\,
\alpha_x^2\lrf{\lambda_u^2}{4\pi}^2|\mathbf{N}_{13}|^2\,
\frac{v_u^2m_{\chi_1^0}^3}{\mu_F^4} 
\label{eq:lspdecay}
\\
&\simeq& (1\times 10^{-6}\,\text{s})^{-1}(N^2\!-1\!)N_F^2|\mathbf{N}_{13}|^2
\lrf{\alpha_x}{10^{-3}}^2\lrf{\lambda_u}{0.75}^4
\lrf{m_{\chi_1^0}}{200\,\gev}^3
\lrf{100\,\tev}{\mu_F}^4 \ ,
\nnmb
\eeq
where $|\mathbf{N}_{13}|$ describes the $\widetilde{H}_d$ content
of the MSSM LSP.
This decay can occur before primordial nucleosynthesis, 
even for very large values of $\mu_F \gtrsim 100\,\tev$.

  Finally, let us mention that the exotic doublets $F$ and $F^c$ will
disrupt standard gauge unification.  This can be restored by embedding
these multiplets in $\mathbf{5}$ and $\mathbf{\bar{5}}$ representations
of $SU(5)$ and limiting the amount of new matter to maintain perturbativity
up to the unification scale~\cite{Feng:2011ik}.  The latter requirement
corresponds to $N\times N_F \leq 5$ for $\mu_F \sim 100\,\tev$.

\subsection{Moduli Reheating and Hidden Dark Matter}

  The treatment of dark matter production by moduli reheating in
this scenario is slightly different from the situations studied
previously.  The key change is that the visible and hidden
sectors are unlikely to reach kinetic equilibrium with one another
after reheating for $\mu_{F,P} \gtrsim m_{3/2}$.  
As a result, it is necessary to keep track of the effective 
visible and hidden temperatures independently. 

  To estimate kinetic equilibration, let us focus on the wino
operator of Eq.~\eqref{wop1}.  This gives rise to $G_x\gamma\to
G_x\gamma$ scattering with a net rate of
$\Gamma \sim T^9/\mu_F^8$.  Comparing to the Hubble rate,
kinetic equilibration requires $\Teq \gtrsim (\mu^8_F/\mpl)^{1/7}$.
On the other hand, the reheating temperature after moduli decay
is on the order $\TRH \sim (\mgrav^3/\mpl)^{1/2}$.
Thus, we see that $\TRH$ is parametrically smaller than
$\Teq$ for $\mu_F\gtrsim m_{3/2}$.  A similar argument applies to
the Higgs interaction in the second term in Eq.~\eqref{wop2}.

  The total modulus decay rate is the sum of partial rates into
the visible and hidden sectors,
\beq
\Gamma_{\varphi} &=& \frac{c}{4\pi}\frac{\mphi^3}{\mpl^2}
~=~ \Gamma_v+\Gamma_x ~=~ 
\frac{c_x+c_v}{4\pi}\frac{\mphi^3}{\mpl^2} \ ,
\eeq
where $c_x$ and $c_v$ describe the relative hidden and visible
decay fractions.  Moduli decays will reheat both sectors independently,
and self-interactions within each sector will lead to self-thermalization.
The total radiation density is the sum of the two sectors,
$\rho_R = \rho_v+\rho_x$.  
We will also define effective temperatures within each sector by
\beq
\rho_v &=&  \frac{\pi^2}{30}g_*T^4 \ ,\\
\rho_x&=& \frac{\pi^2}{30}g_{*x}T_x^4 \ ,
\eeq
where $g_*$ and $T$ refer to the visible sector, and $g_{*x}$ and $T_x$
to the hidden.  Since the hidden and visible sectors 
do not equilibrate with each other after reheating, 
entropy will be conserved independently in both sectors.

Just after reheating, we also have
\beq
\rho_v = \lrf{c_v}{c}\rho_R,~~~~~\rho_x=\lrf{c_x}{c}\rho_R \ .
\eeq
Given the first equality, we now define the reheating temperature
to be
\beq
\TRH = \lrf{c_v}{c}^{1/4}\left[\frac{90}{\pi^2 g_*(\TRH)} \right]^{1/4} 
\sqrt{\Gamma_\varphi \Mpl} \ ,
\label{eq:trhx}
\eeq
corresponding approximately to the visible radiation temperature when 
$H = \Gamma_{\varphi}$.  In the same way, we also define the reheating temperature
in the hidden sector to be
$\TRHx = (c_x/c_v)^{1/4} (g_*/g_{*x})^{1/4}\TRH$.

  The number density of $SU(N)_x$ gaugino dark matter evolves
according to Eq.~\eqref{eq:boltzdm} but with two important modifications.
First, the quantity $\mathcal{N}_{\chi}$ now corresponds to the mean
number of hidden gauginos produced per modulus decay.  This includes
production from direct decays, decay cascades (including decays of the lightest
MSSM neutralino), and re-scattering.  The second key change is that
the thermal average in $\langle\sigma v\rangle$ is now taken over the
hidden-sector distribution with effective temperature $T_x\simeq \TRHx$.

  The thermally-averaged $SU(N)_x$ gaugino cross section  
can receive a non-perturbative
Sommerfeld enhancement from multiple hidden gluon exchange
if the hidden confinement scale is very low, 
as we expect here~\cite{Appelquist:1974zd,Brodsky:1987cv}.
This enhancement can be written as a rescaling of the
perturbative cross section,
\beq
\langle\sigma v\rangle = S_x\langle\sigma v\rangle_{\mathrm{pert}} \ .
\eeq
The perturbative cross section can be obtained by modifying
the $SU(3)_c$ gluino result~~\cite{Baer:1998pg} 
by the appropriate colour factor:
\beq
\langle\sigma v\rangle_{\mathrm{pert}} &=& \frac{3N^2}{16(N^2-1)}\frac{1}{4\pi}
\lrf{g_x^4}{M_x^2} \ .
\eeq
The Sommerfeld enhancement factor is~\cite{Appelquist:1974zd,Brodsky:1987cv,Baer:1998pg}
\beq
S_x = A/(1-e^{-A}) \ ,
\eeq
with $A = \pi\alpha_x/v$, for $v=\sqrt{1-4M_x^2/s}$.  
In the perturbative cross section, the characteristic momentum 
transfer is $\sqrt{s}\simeq 2M_x$, and $\alpha_x$ should be evaluated 
at this scale.  However, the typical momentum transfer leading to 
the non-perturbative enhancement is $\sqrt{s} \sim 2vM_x$~\cite{Baer:1998pg}.  
In our calculation, we estimate $v\simeq \sqrt{3\TRHx/2M_x}$ 
and take $A$ to be
\beq
A \simeq \frac{\pi}{2v}\alpha_x
\left[1
+\frac{11N}{6\pi}\alpha_x\ln(v)\right]^{-1} \ ,
\eeq
where $\alpha_x$ in this expression is evaluated at $2M_x$.

\begin{figure}[ttt]
 \centering
\includegraphics[width=10cm]{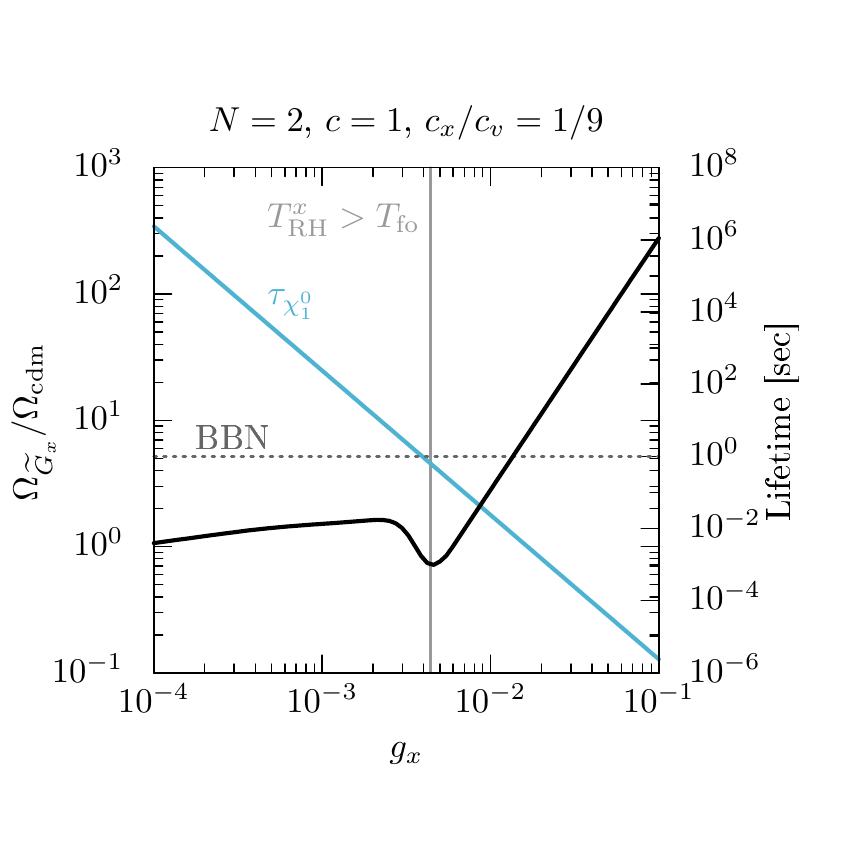}
\vspace{-1cm}
 \caption{Relic abundance of the hidden gluino $\widetilde{G}_x$ (solid black) 
after moduli reheating as a function of the hidden gauge coupling $g_x$
for $N=2$, $\mphi = \mgrav = 100\,\tev$, $c=1$, $\mathcal{N}_x=1$, and $c_x/c_v=1/9$.
The lifetime of the lightest MSSM superpartner, assumed to be a Higgsino-like
neutralino, is shown in light blue for $\mu=150\,\gev$, $N_F=3$, 
and $\lambda_u=0.75$.  The vertical solid grey line corresponds 
to $\TRHx\approx T_{\mathrm{fo}}$, while the dashed horizontal
line shows $\tau_{\chi_1^0} = 1\,\text{s}$.
\label{fig:sun_abun}}
\end{figure}

  In Fig.~\ref{fig:sun_abun} we show the relic density of hidden gluinos 
produced by moduli reheating as a function of $g_x$ for 
$\mphi = m_{3/2} = 100\,\tev$, $c=1$, $\mathcal{N}_x \sim 1$,
and $c_x/c_v = 1/9$.  We also show in this figure the lifetime
of the lightest MSSM superpartner in seconds, which we take to
be a Higgsino-like neutralino with $\mu=150\,\gev$,
along with $N=2$, $\mu_F=\mgrav$, $N_F=3$, and $\lambda_u=0.75$.  
As expected from the estimate of Eq.~\eqref{eq:omscale}, 
smaller values of the gauge coupling $g_x \ll g_2$ are needed 
to obtain an acceptable relic density.

  For very small $g_x$, the hidden gluino mass becomes small enough
that the reheating temperature exceeds the freeze-out temperature,
and the final density is given by the thermal value. This corresponds 
to the plateau, where the abundance is only weakly dependent 
on the gauge coupling. 
At intermediate $g_x$, freeze-out happens in the matter 
dominated phase, where $\Omega_{\widetilde{G}_x}\propto M_x^{-3} \propto g_x^{-6}$~\cite{Giudice:2000ex},
resulting in the turn-over. The abundance continues to decrease until 
non-thermal production takes over, corresponding to the straight section 
for $g_x \gtrsim 4\times 10^{-3}$. 
Note as well that very small values of $g_x$ also increase
the lifetime of the lightest MSSM state to $\tau > 1\,\text{s}$.
This can be problematic for nucleosynthesis, and will be discussed
in more detail below.

\subsection{Hidden Gluino Bounds}

  We found previously that for $M_x < M_2$ and AMSB-like masses,
the $SU(N)_x$ confinement scale is negligibly small relative
to the Hubble scale today.  This implies that the hidden gluon 
will be a new relativistic degree of freedom in
the early Universe.  A nearly massless hidden gluon will also
interact significantly with the relic hidden gluinos, which has
significant implications for dark matter clustering and its imprint
on the CMB.

  New relativistic particles are constrained by primordial
nucleosynthesis and the CMB.  The number of corresponding degrees
of freedom is often written in terms of an effective number 
of additional neutrino species, $\Delta \Neff$.  
If the hidden gluon is the only new light state below 
the reheating temperature and $5\,\mev < \TRH < m_{\mu}$, 
we have~\cite{Feng:2011in}
\beq
\Delta \Neff \simeq \lrf{4}{7}(N^2-1)\lrf{c_x}{c_v} \ ,
\eeq
where $c_x$ and $c_v$ correspond to the hidden and
visible branching fractions of the moduli.
The current upper bound (95\%\,c.l.) on $\Delta \Neff$ 
from primordial nucleosynthesis is~\cite{Cyburt:2004yc,Mangano:2011ar}
\beq
\Delta \Neff \lesssim 
1.0& ~{\rm at}~ &T\sim \TBBN \ .
\eeq
This bound can be satisfied for smaller $N$ provided $(c_x/c_v) < 1$.  
If we reinterpret our moduli results in terms of heavy gravitino decay,
the corresponding ratio is $c_x/c_v = (N^2-1)/12$ if only gaugino modes are open
and $c_x/c_v = 12(N^2-1)/193$ if all MSSM channels 
are available~\cite{Moroi:1995fs}.
A similar limit on $\Neff$ can be derived from 
the CMB~\cite{Valentino:2013wha}.  However, the net effect of the hidden
gluon and gluino on the CMB is more complicated than just a change
in $\Delta \Neff$, as we will discuss below.

  A more significant challenge to this scenario comes from the 
relatively unsuppressed interactions among the hidden gluons 
and gluinos.  Self-interactions among dark matter particles
are strongly constrained by observations of elliptical galaxies and 
the Bullet Cluster~\cite{Ackerman:mha,Feng:2009mn,Buckley:2009in}.\footnote{
Dark matter interactions close to these upper bounds can
help to resolve some of the puzzles of large-scale 
structure~\cite{Boddy:2014yra,Rocha:2012jg,Peter:2012jh,
Vogelsberger:2012ku,Zavala:2012us}.}
Furthermore, we find that the relic hidden gluinos remain kinetically
coupled to the hidden gluon bath until very late times.  This generates
a pressure in the dark gluino fluid that interferes with its gravitational
collapse into bound structures.  A study of this effect lies beyond
the scope of this paper, and we only attempt to describe some of 
the general features here.

  In this scenario, moduli reheating generates a bath of thermal gluons
with temperature $T_x\sim (c_x/c_v)^{1/4}T$. 
Arising from a non-Abelian
gauge group, the gluons will interact with themselves at the rate
\beq
\Gamma \sim \alpha_x^2T_x \sim (10^{-12}\text{eV})\lrf{c_x}{c_v}^{1/4}
\lrf{\alpha_x}{10^{-4}}^2\,\lrf{T}{2.7\,\text{K}} \ .
\eeq
This is easily larger than the Hubble rate today, $H\sim 10^{-33}\,\text{eV}$,
and we expect the hidden gluon to remain in self-equilibrium 
at the present time.  One of the key features of such non-Abelian
plasmas at temperatures well above the confinement scale is that
the gluon field is screened by its 
self-interactions~\cite{Gross:1980br,McLerran:1986zb}.
Correspondingly, the electric and magnetic components of the gluon
develop \emph{Debye masses} on the order of~\cite{Arnold:1995bh},
\beq
m_{E} &\sim& \sqrt{\alpha_x}\;T_x\\
m_{B} &\sim& \alpha_xT_x \ .
\eeq

  Relic hidden gluinos will interact with the hidden gluon bath
through Compton-like scattering.  This can proceed through a $t$-channel
gluon with no suppression by the hidden gluino mass.
Modifying the calculation of Refs.~\cite{Tulin:2013teo}, we find
that the corresponding rate of momentum transfer
between a relic gluino and the gluon bath is much larger than the 
Hubble rate even at the present time.  We also estimate that for moderate
$\alpha_x$ and $\mphi\sim 100\,\tev$ the rate of formation 
of gluino-gluino bound states, which are expected to be
hidden-colour singlets in the ground 
state~\cite{Nanopoulos:1983ys,Kuhn:1983sc,Goldman:1984mj},
is much smaller than the Hubble rate at temperatures below 
the binding energy. 
  
  Together, these two results imply that the relic gluinos remain
kinetically coupled to the gluon bath.  The pressure induced by
the gluons will drive gluinos out of overdense regions and interfere
with structure formation, analogous to the photon pressure felt by baryons
before recombination.  This is very different from the behaviour of
standard collisionless cold dark matter, and implies the hidden
gluinos can only be a small fraction of the total dark matter density.
This fraction, can be constrained using observations
of the CMB and galaxy surveys.  A study along these lines was performed
in Ref.~\cite{Cyr-Racine:2013fsa}, and their results suggest that the fraction
$f_x = \Omega_{\widetilde{G}_x}/\Omega_{\mathrm{cdm}}$ must be less than a few percent,
depending on the temperature ratio $T_x/T \simeq (c_x/c_v)^{1/4}$.\footnote{
A relic population of millicharged particles will have a similar effect.
This was considered in Refs.~\cite{Dubovsky:2003yn,Dolgov:2013una}, 
and a limit of $f_x \lesssim 1\%$ was obtained.} 
Hidden gluino interactions may also modify the distribution of dark matter
on galactic scales~\cite{Fan:2013yva}.

\subsection{Summary}

  This supersymmetric hidden $SU(N)_x$ extension can produce a much smaller
non-thermal LSP relic density than the MSSM, and has only invisible 
annihilation modes that are not constrained by indirect detection.
However, the hidden gluino LSP remains in thermal contact with a
bath of hidden gluons, and thus can only make up at most a few percent
of the total dark matter density.  Obtaining such small relic
densities is non-trivial and leads to new challenges, as we will discuss here.

  From Fig.~\ref{fig:sun_abun} we see that reducing the gauge coupling
$g_x$ lowers the non-thermal hidden gluino density until 
$\TRH \sim T_\text{fo}$,
at which point the relic abundance becomes approximately constant in $g_x$.
At the same time, Eq.~\eqref{eq:lspdecay} shows that smaller values of $g_x$ 
also suppress the decay rate of the lightest MSSM superpartner.
If such decays happen after the onset on primordial nucleosynthesis,
they can disrupt the abundances of light 
elements~\cite{Kawasaki:2004qu,Jedamzik:2006xz}.
The direct two-body decays $\chi_1^0 \to \widetilde{G}_xG_x$ are invisible. 
However, the operator of Eq.~\eqref{wop2} also gives rise to 
the semi-visible three-body mode $\chi_1^0\to h^0\widetilde{G}_xG_x$
if it is kinematically allowed.  The decay products of the Higgs boson 
will be significantly hadronic, and can modify light-element abundances.  
The branching fraction of this three-body mode depends on the available 
phase space.  Taking it to be $B_h \sim 10^{-3}$ and estimating the
Higgsino yield as in Section~\ref{sec:mssm}, we find that Higgsino lifetimes 
below $\tau_{\chi_1^0} \lesssim 1\!-\!100\,\text{s}$ are 
allowed~\cite{Kawasaki:2004qu}.
This can occur for larger values of $N$, $N_F$, or $\lambda_u$,
or smaller values of $\mu$ or $\mu_F$.  Note that reducing $\mu_F$
below $\mphi/2$ is dangerous because it would lead to the production
of stable massive $F$ and $P$ states which would tend to overclose
the Universe.

  An acceptable hidden gluino relic density with a sufficiently
rapid MSSM decay can be obtained in this scenario, but only in
a very restricted and optimistic region of parameters.
For example, with $r_x=3N/5$, $g_x = 0.01$, $N=2$, $N_F=3$, $\lambda_u = 0.75$,
$c_x/c_v = 1/9$, and $\mphi = 2\mgrav = 2\mu_F = 100\,\tev$,
we obtain $\Omega_{\widetilde{G}_x}/\Omega_{\text{cdm}} = 0.023$ 
and $\tau_{\chi_1^0}=0.01\,\text{s}$.  Compared to the parameters
used in Fig.~\ref{fig:sun_abun}, the greatest effect comes from
the small value of $r_x$ relative to the minimal AMSB value ($r_x=3N$).
Such a reduction could arise from threshold corrections due to the
heavy multiplets~\cite{Gupta:2012gu}.

\section{Conclusions\label{sec:conc}}

  In this work we have investigated the production of LSP dark matter
in the wake of moduli oscillation and reheating.  For seemingly generic
string-motivated moduli parameters $\mphi = \mgrav$, $c=1$, 
 $\mathcal{N}_{\chi} \sim 1$, we have argued that the MSSM LSP is typically
created with an abundance that is larger than the observed dark matter
density.  The exception to this is a wino-like LSP, which has been shown
to be inconsistent with current bounds from indirect detection.
We call this the MSSM moduli-induced LSP problem.

  To address this problem, we have studied three gauge extensions of 
the MSSM.  In the first, the MSSM is expanded to include a lighter 
hidden $U(1)_x$ vector multiplet with kinetic mixing with hypercharge that 
is spontaneously broken by a pair of chiral hidden Higgs multiplets.  
The kinetic mixing interaction allows the lightest MSSM superpartner
to decay to the lighter hidden sector LSP.  If this LSP consists 
primarily of the hidden Higgsinos and is sufficiently light, 
it will annihilate very efficiently.
The resulting hidden LSP relic abundance after moduli reheating
can be small enough to be consistent with current bounds from 
indirect detection and the CMB.  In this case, a second more abundant
component of the DM density is needed.  The spectrum of scalar soft terms
required in this theory can also be challenging to obtain for
the large values of $\mgrav \gtrsim 100\,\tev$ considered.

  The second extension of the MSSM that we studied has an
asymmetric dark matter candidate.  The underlying theory in this
case was again a kinetically-mixed $U(1)_x$ vector multiplet
spontaneously broken by a pair of chiral hidden Higgs, but now
with an additional pair of chiral multiplets $Y$ and $Y^c$. 
For a range of parameters, the two stable states in this theory
are the Dirac fermion $\Psi$ and the lighter complex scalar $\Phi$
derived from $Y$ and $Y^c$.  If $\Psi$ or $\Phi$ obtain a significant
particle anti-particle asymmetry in the course of moduli reheating,
they can account for the entire DM density.  A large production asymmetry
leads to a very small residual anti-DM component, which allows the
asymmetric abundances of $\Psi$ and $\Phi$ to be consistent with 
limits from indirect (and direct) detection.  However, the production
asymmetry required for this to work is relatively large, and may be
difficult to obtain in a more complete theory of asymmetry generation.
This theory also faces the same scalar soft term requirement as
the symmetric hidden $U(1)_x$ extension.

  The third extension of the MSSM consists of a pure
non-Abelian $SU(N)_x$ vector multiplet at low energies.  This sector 
can connect to the MSSM through additional heavy multiplets
charged under both the visible and hidden gauge groups,
allowing for decays of the lightest MSSM superpartner to the $SU(N)_x$ gluino.
Acceptable hidden gluino relic densities can be obtained for smaller
values of the $SU(N)_x$ gauge coupling.  This implies a potential
tension with primordial nucleosynthesis from late MSSM decays,
and leads to a negligibly small hidden confinement scale.
In contrast to the two previous extensions, light scalar superpartners
are not required and this mechanism can work in the context of mini-split
supersymmetry~\cite{Wells:2004di,Hall:2011jd,
Arvanitaki:2012ps,ArkaniHamed:2012gw,Ibe:2011aa,Ibe:2012hu}.
While this theory is not constrained by standard indirect detection searches,
the coupling of the hidden gluino to a bath of hidden gluons leads to
non-standard DM dynamics that require the hidden gluino density to
be only a few percent of the total DM density.  It is very difficult
to obtain relic densities this small in this scenario.

  Our main conclusion is that it is challenging to avoid producing
too much LSP dark matter in the course of string-motivated moduli reheating.  
For seemingly generic modulus parameters, the relic density in the 
MSSM is either too large or at odds with limits from 
indirect detection.  This may be a hint that the properties of
moduli (in our vacuum at least) differ from the general expectations
discussed above~\cite{Acharya:2012tw,Bose:2013fqa}.  
Alternatively, this could be an indication of new light physics 
beyond the MSSM.  
We have considered three examples of the latter possibility in this paper
and have shown that they can produce a stable LSP
abundance that is consistent with current observations and limits.
Even so, these three extensions all lead to a significant complication
of the MSSM and require a somewhat fortuitous conspiracy of parameters
for them to succeed.  A more direct solution might be the absence of
a stable LSP through $R$-parity violation, or simply the absence
of light superpartners and very large $\mphi\sim \mgrav$.

\section*{Acknowledgements}

  We thank Matthew Reece, Kris Sigurdson, Scott Watson, and Kathryn Zurek
for helpful discussions.  This work is supported by the Natural Sciences
and Engineering Research Council of Canada~(NSERC).


\end{document}